\documentclass[letterpaper,twocolumn,10pt]{article}
\usepackage{usenix-2020-09}

\usepackage{tikz}
\usepackage{amsmath,amsthm}

\usepackage{tabularx}
\usepackage{subcaption}
\usepackage{algorithm}
\usepackage[noend]{algorithmic}
\usepackage{titling}

\usepackage{xspace}
\usepackage[textsize=scriptsize]{todonotes}
\usepackage{listings}
\usepackage{color}

\definecolor{dkgreen}{rgb}{0,0.6,0}
\definecolor{gray}{rgb}{0.5,0.5,0.5}
\definecolor{mauve}{rgb}{0.58,0,0.82}

\newcommand{\sysname}{\textsc{TACCL}\xspace}
\newcommand{\vheading}[1]{\vspace{0.0125in}\noindent\textbf{#1}}

\newcommand{\allgather}{\textsc{Allgather}\xspace}

\newcommand{\reducescatter}{\textsc{Reducescatter}\xspace}
\newcommand{\allreduce}{\textsc{Allreduce}\xspace}
\newcommand{\alltoall}{\textsc{Alltoall}\xspace}

\newcommand{\dgxtwo}{DGX-2\xspace}

\newcommand{\ndvtwo}{NDv2\xspace}
\newcommand{\ucmax}{\texttt{uc-max}\xspace}
\newcommand{\ucmin}{\texttt{uc-min}\xspace}

\newcommand{\camera}[1]{#1}
\newcolumntype{x}[1]{>{\centering\arraybackslash}p{#1pt}}
\newcommand{\tablestyle}[2]{\setlength{\tabcolsep}{#1}\renewcommand{\arraystretch}{#2}\centering\footnotesize}
\newlength\savewidth\newcommand\shline{\noalign{\global\savewidth\arrayrulewidth
  \global\arrayrulewidth 1pt}\hline\noalign{\global\arrayrulewidth\savewidth}}

\providecommand{\ie}{i.e., }
\providecommand{\eg}{e.g., }
\providecommand{\myparab}[1]{\smallskip\noindent\textbf{#1} }
\newcolumntype{P}[1]{>{\centering\arraybackslash}p{#1}}
\newcolumntype{M}[1]{>{\centering\arraybackslash}m{#1}}

\newenvironment{packeditemize}{\begin{list}{$\bullet$}{\setlength{\itemsep}{0.2pt}\addtolength{\labelwidth}{0pt}\setlength{\leftmargin}{\labelwidth}\setlength{\listparindent}{\parindent}\setlength{\parsep}{1pt}\setlength{\topsep}{0pt}}}{\end{list}}

\theoremstyle{definition}
\newtheorem{example}{Example}[section]
\begin{document}

\date{}

\setlength{\droptitle}{-4em}   
\posttitle{\par\end{center}}

\title{\Large \bf \sysname: Guiding Collective Algorithm Synthesis using Communication Sketches}
\author{
{\rm Aashaka Shah}\thanks{Work was partially done during an internship at Microsoft Research.}\\
University of Texas at Austin
\and
{\rm Vijay Chidambaram}\\
University of Texas at Austin and VMware Research
\and
{\rm Meghan Cowan}\\
Microsoft Research
\and
{\rm Saeed Maleki}\\
Microsoft Research
\and
{\rm Madan Musuvathi}\\
Microsoft Research
\and
{\rm Todd Mytkowicz}\\
Microsoft Research
\and
{\rm Jacob Nelson}\\
Microsoft Research
\and
{\rm Olli Saarikivi}\\
Microsoft Research
\and
{\rm Rachee Singh}\\
Microsoft and Cornell University
} 
\maketitle

\begin{abstract}
Machine learning models are increasingly being trained across multiple GPUs and servers.
In this setting, data is transferred between GPUs using communication collectives such as \alltoall and \allreduce,
which can become a significant bottleneck in training large models. Thus, it is important to use efficient algorithms for collective communication.
We develop \sysname, a tool that enables algorithm designers to guide a synthesizer into automatically generating algorithms for a given hardware configuration and communication collective.
\sysname uses a novel \emph{communication sketch} abstraction to get crucial information from the designer to significantly reduce the search space and guide the synthesizer towards better algorithms.
\sysname also uses a novel encoding of the problem that allows it to scale beyond single-node topologies.
We use \sysname to synthesize algorithms for three collectives and two hardware topologies: DGX-2 and NDv2.
We demonstrate that the algorithms synthesized by \sysname outperform the Nvidia Collective Communication Library (NCCL) by up to 6.7$\times$.
We also show that \sysname can speed up end-to-end training of Transformer-XL and BERT models by 11\%--2.3$\times$ for different batch sizes.

\end{abstract}

\section{Introduction}

Machine-learning models have been dramatically increasing in size over the past few years.
For example, the language model MT-NLG has
530 billion parameters~\cite{mtnlg} and the Switch-C mixture-of-experts model has 1.6 trillion parameters~\cite{switch}.
Model sizes are expected to further grow to increase model accuracy
and perform more complex tasks.
These models are too large for the resources of a single GPU and have to be
distributed across multiple servers, each with several GPUs,
using different parallelism strategies like data, model, pipeline, and expert
parallelism~\cite{megatron,gshard,switch} for training and inference.
Intermediate data and
parameters of the model at each GPU are accumulated, shuffled, and transferred
over the network between other GPUs for distributed machine learning,
depending on the type of parallelism strategy used.

\myparab{The inter-GPU communication bottleneck.}
Recent work has shown that GPU idle time spent waiting for network communication
can be significant in practice~\cite{switchML2021,ATP2021,panama2021,phub2018}.
For instance, BERT~\cite{bert} and
DeepLight~\cite{deeplight} spent 11\% and 63\% of time, respectively,
with GPUs idle on a 100 Gbps Ethernet cluster of P100 GPUs~\cite{switchML2021}.
Newer generations of faster GPUs
will only make this problem worse. This inefficient use of
GPUs shows that there is significant model performance to be
gained by optimizing inter-GPU communication. 

\myparab{Collective communication primitives and algorithms.}
Efficient communication between GPUs is the key to enabling fast distributed
ML training and inference. Modern GPU systems use message passing interface 
(MPI)-based \emph{collective communication primitives}, such as \allreduce, 
\allgather, and \alltoall to perform inter-GPU communication (Figure~\ref{fig:collectives} in \S\ref{sec:background}).
\emph{Collective algorithms} implement collective communication primitives.
They route data along various paths in the network and schedule the
necessary computation (\eg a sum in \allreduce) while optimizing for latency and bandwidth characteristics of each
link in the network. For example, a common collective algorithm for \allgather (all GPUs gather data from all GPUs)
is a Ring algorithm, in which all GPUs are logically arranged in a ring and each GPU receives data from its predecessor
in the ring and sends a previously received data to its successor.
Inefficiencies in collective communication algorithms can cause poor
network utilization, causing GPUs to remain
idle until inter-GPU transfers complete~\cite{zhang2020network}, and thus reducing the overall
efficiency of distributed training and inference.

\begin{figure*}[t]
    \centering
    \includegraphics[width=0.8\linewidth]{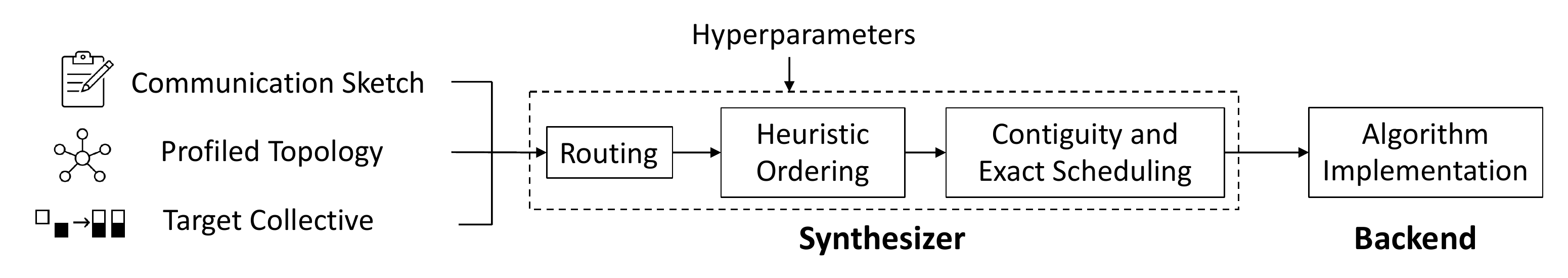}\\
    \caption{\sysname's novel synthesizer takes as input a communication sketch,
    profiled topology, and target collective along with synthesizer hyperparameters
    to generate an algorithm for the collective. The synthesized algorithm is
    implemented on the hardware cluster using \sysname's backend.}
    \label{fig:workflow}
\end{figure*}

\myparab{Challenges in designing GPU communication algorithms.}
Designing algorithms for efficient collective communication on GPU
topologies is challenging. First, these algorithms have to strike the right
balance between latency and bandwidth optimality.
For instance, the commonly used Ring algorithm for \allreduce
is not efficient for small input sizes as it has a high latency.
Second, GPU communication algorithms have to manage the heterogeneity of
connectivity in the underlying topology. For instance, GPUs within a machine (also referred to as a node) are usually 
connected using fast NVLinks~\cite{nvlink} (up to 300 GBps aggregate bidirectional bandwidth per GPU)
while GPUs across nodes are connected using slow InfiniBand~\cite{infiniband} links
(12.5-25 GBps per NIC). Moreover, these topologies vary significantly between 
vendors. And finally, searching over the entire space of routing and scheduling 
algorithms to find optimal ones for communication collectives
is computationally prohibitive.  In fact, previous approaches that synthesize 
collective communication algorithms are limited to single-node
topologies~\cite{cai2021synthesizing} or 8 GPUs at most~\cite{wang2018blink}.

\myparab{Managing scale for automated algorithm design.}
Our goal is to automatically obtain efficient algorithms for a given 
hardware configuration and communication collective.
We encode the problem of finding optimal algorithms for communication
collectives into a mixed integer linear program (MILP) with the goal
of minimizing the overall execution time. Unfortunately, this problem is NP-hard; 
state-of-the-art commercial solvers like Gurobi~\cite{gurobi}
can spend several days exploring the search space without finding
an optimal algorithm. In this work, we propose a \emph{human-in-the-loop} approach 
that incorporates high-level inputs of an algorithm designer
to efficiently synthesize collective communication algorithms for
heterogeneous GPU topologies. We argue that it is easy for 
algorithm designers to provide a few simple inputs that constrain 
the search space of algorithms which allows synthesis engines
to scale to large GPU topologies.

\myparab{Communication sketches as user input.}
It is crucial that the input required from algorithm designers 
is simple and intuitive. For this, we introduce a new abstraction: \emph{communication sketches} (\S\ref{sec:sketches}).
Inspired by the technique of \emph{program sketching}~\cite{solarlezama2008}
from program synthesis, in which developers supply a partially specified program
with holes that capture the high level structure of the desired program,
communication sketches allow algorithm designers to provide high-level 
intuitions that constrain the search space of algorithms. A
synthesis engine fills in the remaining details such as routing and scheduling
of the final collective communication algorithm, analogous to how a constraint
solver in program synthesis searches the reduced space to fill the holes.

\myparab{Our solution.}
We develop \sysname (Topology Aware Collective Communication Library), a system 
that synthesizes communication algorithms for a given topology and a target collective 
communication primitive. Algorithm designers can use communication sketches to guide \sysname into
synthesizing efficient algorithms for a large range of hardware topologies.
We develop a novel encoding of the problem in \sysname's synthesizer to scale 
beyond single-node topologies. 
Figure~\ref{fig:workflow} shows an overview of \sysname's design.

\myparab{Synthesizing algorithms from communication sketches.}
\sysname's synthesis approach
builds on the solver based synthesis approach in SCCL~\cite{cai2021synthesizing},
where the space of possible algorithms is directly encoded in a
\emph{satisfiability modulo-theories} (SMT) solver.
SCCL does not scale to the sizes of clusters used by modern machine learning
workloads. We present a novel \emph{mixed integer linear programming} (MILP) encoding of
the collective algorithm synthesis problem that improves scalability by first solving a
bandwidth-relaxed version of the problem to decide on \emph{routing}, followed by ordering
heuristics and a second
bandwidth-constrained problem to find a valid \emph{scheduling} (\S\ref{sec:distsccl-synthesizer}). In
addition to improving scalability, \sysname's MILP formulation allows modeling of
heterogeneous links with different per-message overhead characteristics. This overcomes the
limitation in SCCL~\cite{cai2021synthesizing} that
prevents it from faithfully targeting distributed GPU clusters.

\myparab{Results.}
We use \sysname to synthesize efficient algorithms for a range of 
collectives like \allgather, \alltoall, and \allreduce, and for different 
hardware backends like Azure \ndvtwo~\cite{ndv2-azure} and Nvidia 
\dgxtwo~\cite{dgx} (\S\ref{sec:eval}).
We compare \sysname to the state-of-the-art Nvidia 
Collective Communication Library (NCCL).
\sysname synthesized an \allgather algorithm for two Nvidia \dgxtwo nodes (32 GPUs).
This algorithm is up-to \textbf{$6.7\times$} faster than 
NCCL for small-to-moderate input sizes.
For large input sizes on the same 
hardware, \sysname synthesized a different \allgather algorithm 
that nearly saturates the inter-node bandwidth and is up-to $25\%$ faster
than NCCL. \sysname synthesized an \alltoall algorithm for two Azure
\ndvtwo nodes (16 GPUs) that is up-to $66\%$ faster than NCCL. 
Finally, we replaced NCCL with \sysname using only 
a two-line code change in PyTorch and found that \sysname
achieves a speed-up of $17\%$ in end-to-end training of a 
mixture-of-experts model that uses \alltoall and \allreduce, 
and a speed-up of $11\%$ - $2\times$ in end-to-end training
of a Transformer-XL model distributed over 16 GPUs for varying batch sizes.
\sysname's codebase is open-source and is actively in use by researchers at
universities and practitioners at Microsoft for Azure's GPU 
virtual machines.~\footnote{\url{https://github.com/microsoft/taccl}}

\section{Background and Motivation}
\label{sec:background}

\begin{figure*}[tb]
    \centering
    \includegraphics[width=\linewidth]{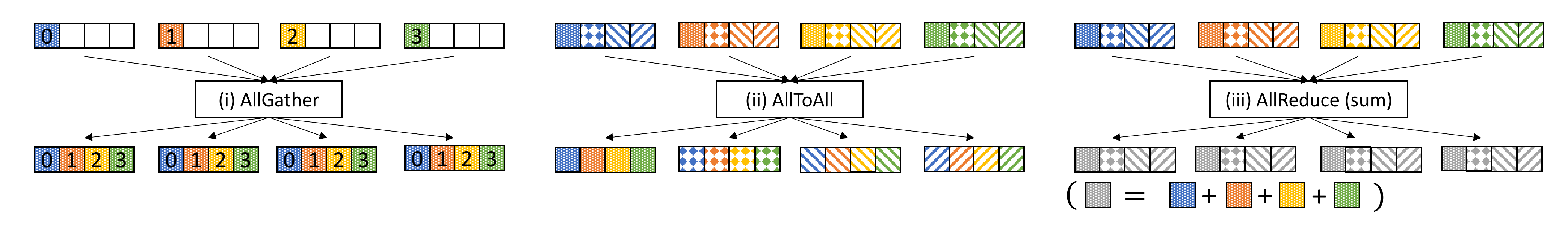}
    \caption{The initial and final data buffers on four GPUs participating in different collectives.}
    \label{fig:collectives}
  \end{figure*}

\myparab{Collective communication in distributed ML workloads.}
Multi-GPU ML workloads typically communicate using MPI-style collectives
like \allgather, \alltoall, and \allreduce shown in Figure~\ref{fig:collectives}. These primitives
capture the application's intent behind the communication,
thus allowing collective communication libraries to optimize for specific hardware configurations.
In \allgather, every GPU receives the data buffers of all other GPUs (left diagram in Figure~\ref{fig:collectives}).
In \alltoall, every GPU receives different parts, or chunks, of the
data buffers present on all GPUs.
This effectively transposes the data chunk from buffer index to GPU index as
can be seen in center diagram in Figure~\ref{fig:collectives}.
In \allreduce, every GPU ends up with a data buffer that has the results
of performing a point-wise computation (\eg sum in right diagram in Figure~\ref{fig:collectives})
over the same data index of all GPUs.

The parallelism strategy for the distributed ML workload determines
which collective communication primitive is used.
Data parallelism and some tensor model parallelisms~\cite{megatron} make use of the
\allreduce collective to aggregate gradients and intermediate data respectively from
multiple GPUs.
Expert parallelism~\cite{gshard,switch} and common deep learning recommendation models
(DLRM)~\cite{mudigere2022software} make use of the \alltoall collective to shuffle
intermediate data between experts and embedding lookup data between GPUs respectively.
DLRMs~\cite{mudigere2022software} also make use of the \allgather collective and another
\reducescatter collective to perform embedding lookups from embedding tables sharded over
multiple GPUs.

\myparab{Existing approaches to collective algorithms.}
Collective algorithms must be designed considering the
target input sizes and the heterogeneity of the target topology.
However, most collective communication libraries used for distributed ML today,
including the state-of-the-art NCCL, use pre-defined templates
of collective algorithms superimposed onto a target topology.
For example, for collectives like \allgather and \reducescatter,
NCCL identifies rings in the target topology and uses the Ring algorithm. For $n$ GPUs, this algorithm requires
$n-1$ link transfer steps per data chunk and is not ideal for
smaller data sizes where link transfer latencies dominate. Further, this
algorithm treats the slow inter-node and fast intra-node links similarly,
scheduling equal number of data transfers across both. The communication is thus bottlenecked
on the slower inter-node links, when it could have benefitted by sending more
node-local data (i.e. data of GPUs local to the node) over the faster intra-node links instead.

For the \alltoall collective, NCCL implements the collective algorithm as
peer-to-peer data transfers between all pairs of GPUs. This algorithm is topology-agnostic
and  often inefficient.
For the \allreduce collective, NCCL chooses between two algorithms --- Double-Binary-Tree~\cite{nccl-tree}
and Ring. This decision is made according to the communication input size and number of nodes, but might not
be most accurate, as it is
based on hardcoded latency and bandwidth profiling done previously by Nvidia on their machines.

Designing efficient collective algorithms requires careful analysis of the topology and its performance with 
different buffer sizes.
Recent work~\cite{wang2018blink,cai2021synthesizing} has shown that synthesis
is a promising approach for generating collective algorithms for
different topologies and to achieve bandwidth and latency optimality.
However, scaling these approaches to multi-node (i.e. multi-machine) distributed GPU topologies has been a challenge.
We measured the synthesis time for \allgather and \alltoall collectives on topologies of two 
Azure NDv2 nodes and two Nvidia DGX2 nodes (Figure~\ref{fig:topos}) using SCCL~\cite{cai2021synthesizing,scclrepo}.
We modified the codebase to include both topologies and attempted to synthesize the collectives
with a 24-hour time limit set for each synthesis query. Given a 24-hour time limit, 
SCCL's \texttt{pareto-optimal} solver strategy
did not finish synthesis for any combination of collective and topology.
The only algorithm that SCCL could synthesize within the time limit was
a latency optimal algorithm for \allgather on two NDv2 nodes.

\myparab{Low-effort inputs from algorithm designers.}
The search space of possible algorithms to implement a collective is intractably large
and cannot be explored via brute-force.
Deciding whether or not to route data chunks from $n$ GPUs over
$l$ links in a topology has $O(2^{n \times l})$ combinations.
As we scale to multi-node topologies, $n$ as well as $l$ will also scale, increasing
the exponent quadratically.
The search space explodes further if we consider the problem of ordering data sends
at each link along with deciding routing for the data.
We argue that high-level inputs from a human algorithm designer 
help reduce the search space to make algorithm synthesis more tractable.
In the most extreme case, the designer would hand-write the entire
algorithm. However, handcrafting data routing and scheduling over links to
implement a collective is complex and requires many design choices.
Instead, designers only provide input in the form of a
communication sketch around which \sysname synthesizes an algorithm.
Our goal is to ensure that providing inputs is a low-effort activity,
but can discard large parts of the search space to
achieve improvements in running-time of the synthesis engine.

\myparab{Synthesis technique.}
\sysname synthesizes a collective algorithm by deciding the route that each data
chunk in the collective should take in the topology as well as the ordering of
chunks at every link.
Even with communication sketches which reduces the search space for the synthesizer,
this decision problem is NP-hard and the complexity increases exponentially with number of GPUs.
To make the problem more tractable, we first relax the synthesis problem to solve just the
routing of all data chunks and then
heuristically order chunks sent over the same links according to bandwidth constraints.
\sysname's synthesizer design along with communication sketches help \sysname
synthesize efficient collectives for multi-node topologies.
\section{Communication Sketches}
\label{sec:sketches}

This paper proposes a new form of sketching~\cite{solarlezama2008} as an effective tool for users to communicate
interesting aspects of collective communication algorithms to synthesis backends. Sketching approaches must strike a
balance between allowing users to omit implementation details while still providing enough direction for the synthesis
to scale. In our experience, \emph{routing} is an aspect of collective communication that we often have intuitions
about, while reasoning about \emph{scheduling} tends to be tedious and better left to synthesis. Moreover, properties
about scheduling are routing dependent since the order of operations is only relevant when routes intersect, which makes
them harder to express. Meanwhile, interesting properties about routing are expressible globally, e.g., ``never send
over the InfiniBand NIC from this GPU''.
Therefore, we ask the algorithm designer (user) for four low-effort inputs as a part of the communication
sketch:
\begin{packeditemize}
    \item Specify the \emph{logical topology} as a subset of the 
    actual physical topology that the algorithm will operate on.
    This constrains the routes chosen by the communication algorithm
    and alleviates over-subscription of low-bandwidth links. 
    For example, the
    outgoing links of all but one GPU can be removed in the logical topology
    to force all data going to remote GPUs to be relayed through one GPU.
    \item The logical topology abstracts away switches (\eg NVSwitches, IBSwitches)
    in the GPU network.
    Users can annotate switches in the topology for the synthesizer
    to use certain \emph{switch-hyperedge policies}, enabling it to apply
    synthesis policies that help algorithms avoid contention.
    \item Provide \emph{algorithm symmetry} based
    on the symmetries in the topology and the collective.
    \item Specify the expected \emph{input size} of the data,
    which is used as a part of the synthesis engine's built-in cost model.
\end{packeditemize}
We explain all parts of the communication sketch and provide
an example sketch written for \sysname in Appendix~\ref{app:sketch}.

\subsection{Logical Topology}
\label{subsec:logicaltopo}

The core of \sysname's communication sketch is a \emph{logical topology}
consisting of a subset of the physical topology after excluding links
that the user prefers \sysname to avoid. A logical topology has as
many nodes as the physical topology and inherits the cost
model produced by \sysname's profiler for the physical topology. 
Logical topologies omit NICs and switches and use switch-hyperedges 
(Section~\ref{subsec:switch-hyperedges}), abstracting them away into links between GPUs. 
The reason is two-fold: \sysname runtime is embedded in NCCL
runtime and NCCL has no direct control over NIC or switch use, and it 
allows \sysname to reason over a smaller graph thus enhancing scalability. 
Section~\ref{subsec:switch-hyperedges} discusses implications of this
abstraction.

\begin{example}[Sketching inside an \ndvtwo]
    Consider the physical topology of an Azure \ndvtwo, given by the union of Figure~\ref{fig:ndv2} and
    Figure~\ref{fig:ndv2-physical}. While NCCL is able to communicate over both the NVLink and PCIe connections, the
    bandwidth offered by the NVLinks is much higher than that of PCIe, and thus it is reasonable to set the logical
    topology to just the NVLink subgraph in Figure~\ref{fig:ndv2}.
\end{example}

\begin{example}[Distributed sketching for \ndvtwo clusters]
    \label{ex:dist-ndvtwo}
    It is essential to use PCIe connectivity for distributed collective 
    communication with multiple \ndvtwo systems since the NIC is connected 
    to GPUs over PCIe (Figure~\ref{fig:ndv2-physical}). Due to lack of GPUDirect 
    RDMA~\cite{gpudirect-rdma} on these systems, all communication over 
    PCIe must pass through host memory. Therefore, care must be taken in 
    choosing which links to use, as the PCIe links between PCIe switches 
    and the CPU are oversubscribed. Obtaining maximum throughput communication
    requires a logical topology that avoids conflicting flows on the oversubscribed PCIe links. To build a logical
    topology for a cluster of \ndvtwo systems, a pair of receiver and sender GPUs is selected for each \ndvtwo such
    that the selected GPUs and the NIC are connected to the same PCIe switch.
\end{example}

\subsection{Switch-Hyperedges}
\label{subsec:switch-hyperedges}

\begin{figure}[t]
    \centering
    \begin{subfigure}[b]{0.56\columnwidth}
        \caption{\small{Physical topology with a switch.}}
        \includegraphics[width=\columnwidth]{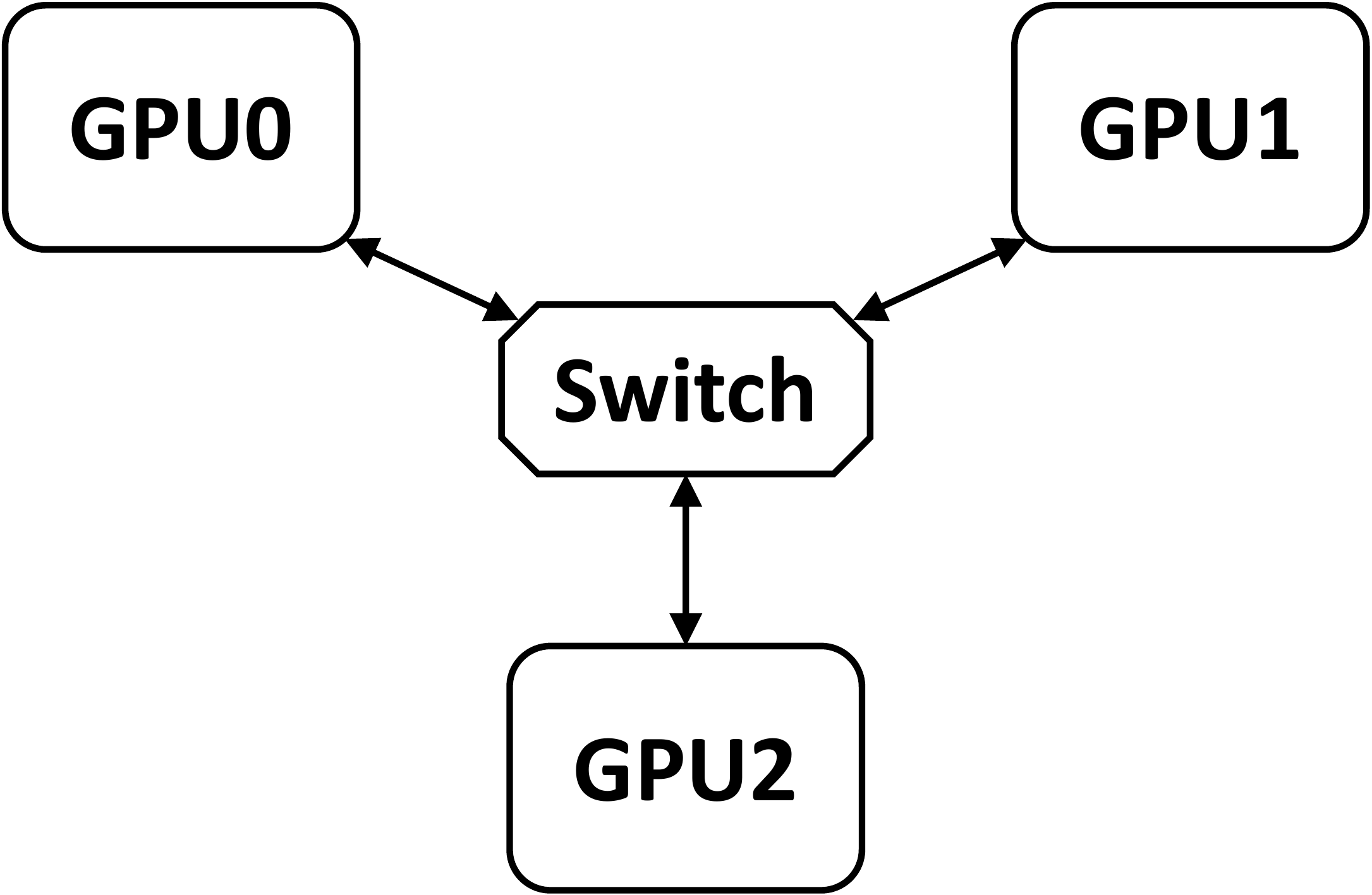}
        \label{fig:switch-physical}
    \end{subfigure}
    \\
    \begin{subfigure}[b]{0.48\columnwidth}
        \includegraphics[width=\columnwidth]{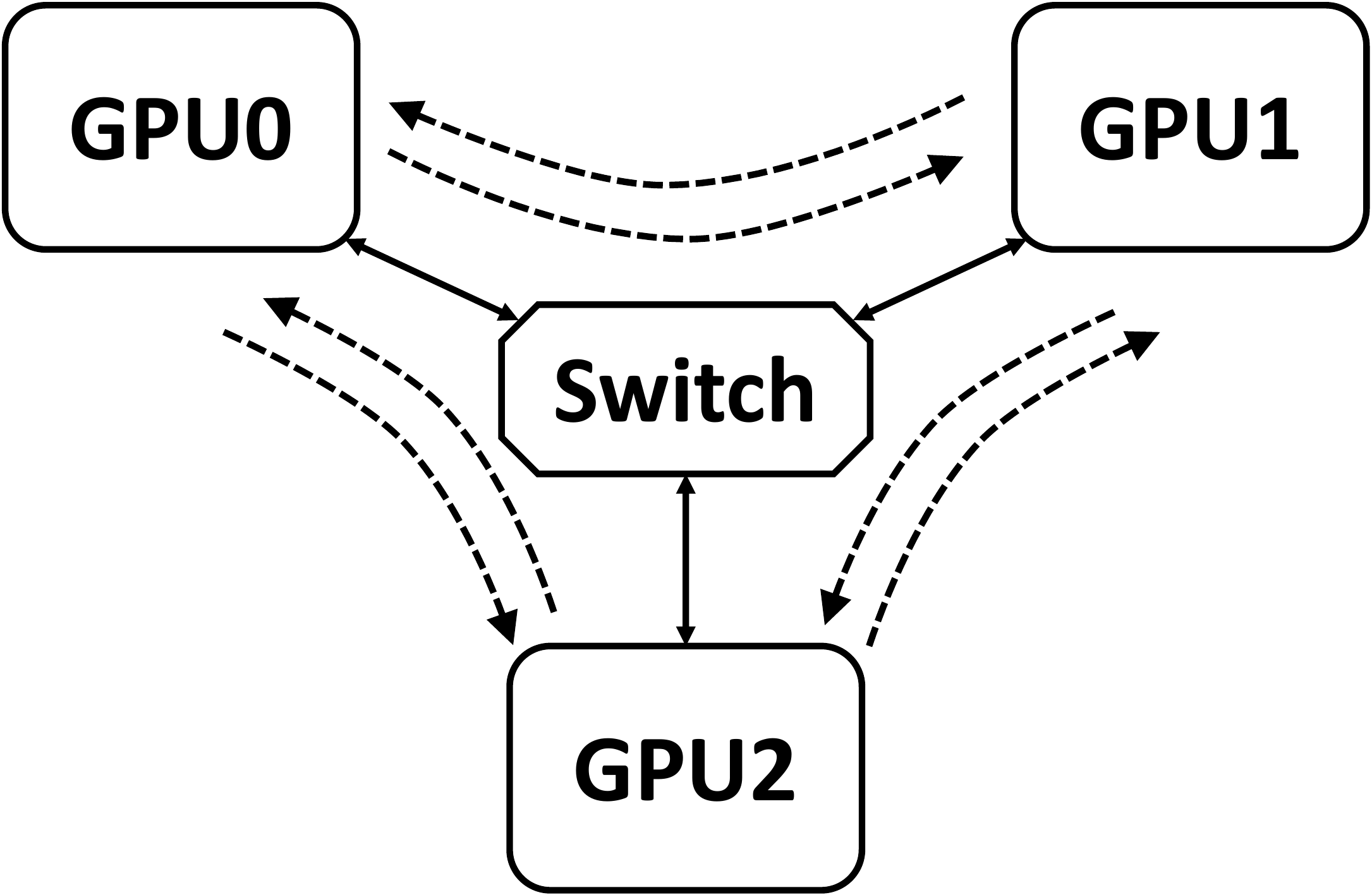}
        \caption{\small{Max. connections strategy.}}
        \label{fig:switch-uc-max}
    \end{subfigure}
    \begin{subfigure}[b]{0.48\columnwidth}
        \includegraphics[width=\columnwidth]{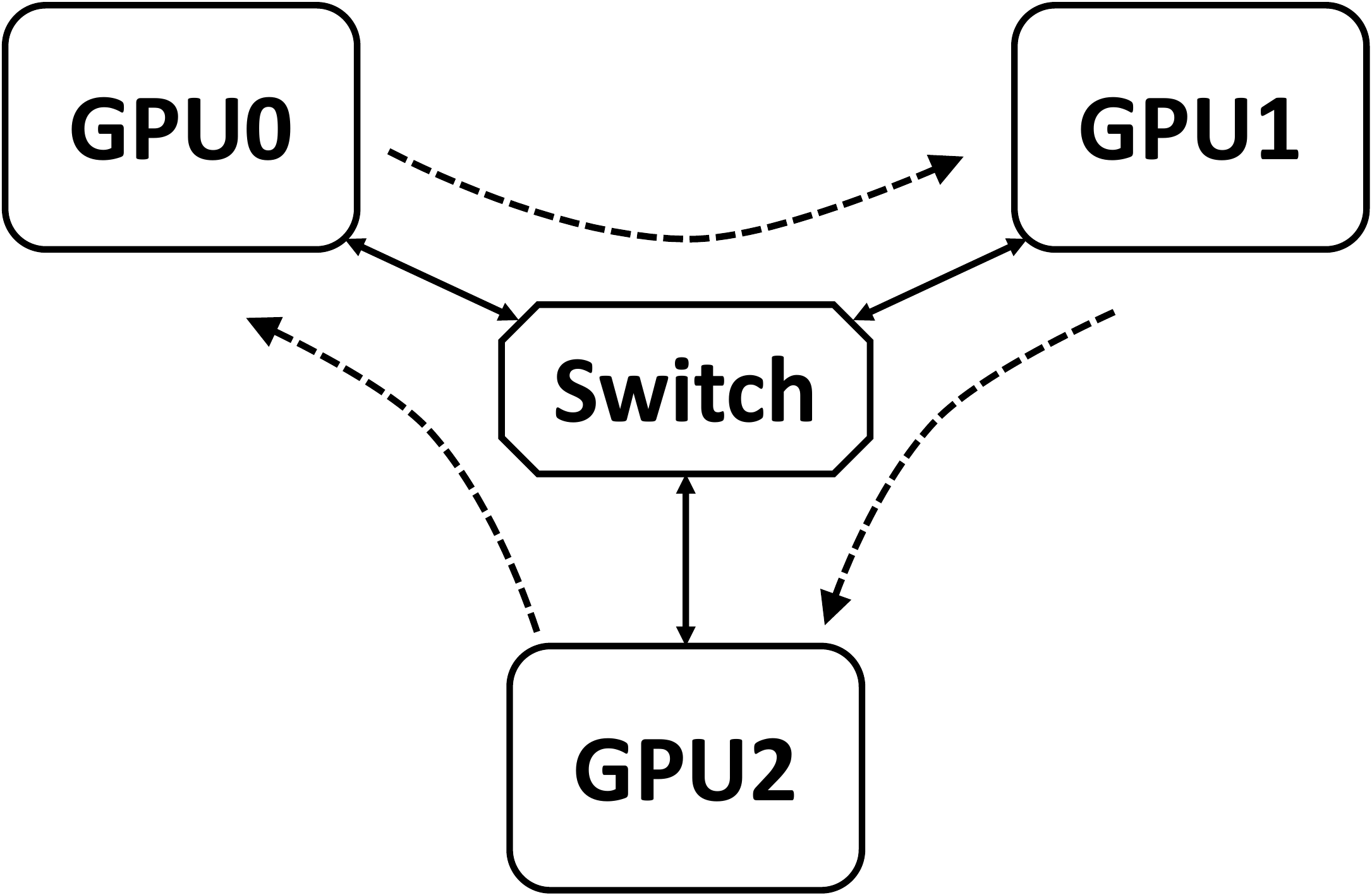}
        \caption{\small{Min. connections strategy.}}
        \label{fig:switch-uc-min}
    \end{subfigure}
    \caption{\small{Effects of switch-hyperedge policies.}}
\end{figure}

\begin{figure*}[t]
    \centering
    \includegraphics[width=\textwidth]{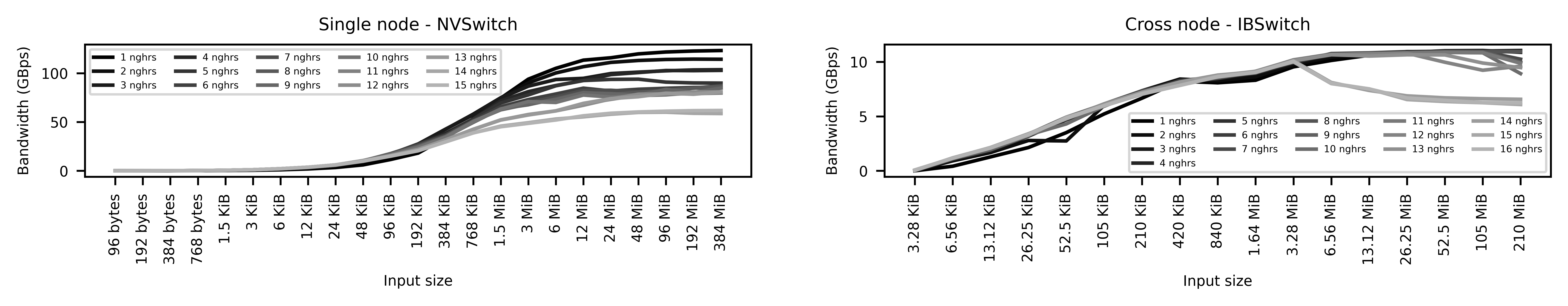}
    \caption{Multi-connection with varying number of GPU neighbors and data volume.}
    \label{fig:multiconnection}
\end{figure*}

In a switched fabric with full bisectional-bandwidth, like the NVSwitch or IBSwitch fabrics in DGX-2 and \ndvtwo
systems, nodes can simultaneously communicate at the full bandwidth of their ingress or egress links. However, as the
number of connections through a switch, originating from a single GPU or NIC increases, the resulting queuing delays increase
the latency. 
Figure~\ref{fig:multiconnection} plots the accumulated ingress/egress 
bandwidth of exchanging varying volume of data (up-to 200-400 MB) for 
different number of connections over NVSwitches in a DGX2 node (left) and 
over IBSwitches among four DGX2 nodes (right).
In both cases, the bandwidth drops as the number of connections increases 
despite the volume of data remaining constant. However,
for small input sizes, the difference for different number of connections 
is not significant. 
\sysname's logical topology does not model switches and does not
capture the effect of number of connections.

\sysname incorporates the effect of multiple connections
using \emph{switch-hyperedges} in the synthesizer to control
the number of connections between GPUs and switches. A switch-hyperedge 
replaces a switch with a set of direct links in the logical topology
for the entire runtime of an algorithm. The synthesizer 
still has the freedom to select which direct links are imposed. To control 
the number of direct links for each switch-hyperedge,
\sysname provides three policies for a user: (1) \emph{maximize} the number of 
links, (2) \emph{minimize} the number of links, and (3)
freely choose any number of links. These policies are enforced by adding
the number of Connections to the objective function (see Appendix~\ref{app:pathenc}
for details).

\begin{example}[Sketching for congestion]
    Figure~\ref{fig:switch-physical} shows a physical topology of three GPUs connected by a switch, where each GPU can
    communicate with any other GPU. 

    Figure~\ref{fig:switch-uc-max} shows a
    logical topology with a switch-hyperedge that \sysname may choose with maximizing number of connections policy. 
    This is desirable for small data sizes that result in low likelihood of congestion at the switch with large number of connections as shown in Figure~\ref{fig:multiconnection}.

    In Figure~\ref{fig:switch-uc-min} \sysname has minimized the number of connections, effectively resulting in a Ring topology.
    This is desirable for larger data sizes, as restricting the number of logical connections limits the congestion in the
    switch (Figure~\ref{fig:multiconnection}).
\end{example}

\subsection{Algorithm Symmetry}

Many collective communication algorithms are symmetric in a variety of ways. 
For example, ring algorithms follow a ring symmetry or in hierarchical 
algorithms, the local phases inside all machines are symmetric to each other. 
Inspired by this, \sysname offers a generic way to enforce algorithms 
to be symmetric.

The user may enforce a symmetry by supplying an \emph{automorphism} of the logical topology and collective, i.e., a
permutation of the ranks and chunks that maintains the structure of the topology and the collective pre- and
post-conditions, and a \emph{partition} of the ranks such that the automorphism maps each subset of ranks to some subset
of the partition. \sysname will then restrict synthesis to algorithms with the same symmetry for all chunk transfers.

\begin{example}
    \label{ex:ndv2-symmetry}
    Consider a cluster of two \ndvtwo systems and the task of synthesizing an \allgather. A hierarchical symmetry may be
    specified with an automorphism composed of a the permutation $[8,\dotsc,15,0,\dotsc,7]$ for both chunks and ranks,
    and a partition $\{\{0,\dotsc,7\},\{8,\dotsc,15\}\}$. Now if an algorithm performs a send of chunk 0 from rank 0 to
    rank 1, then it must also include a send of chunk 8 from rank 8 to rank 9. However, sends between GPUs in different
    \ndvtwo{}s, e.g., between 0 and 8, are not affected by the symmetry.

    Since the internal topologies of \ndvtwo systems are identical, enforcing this symmetry is reasonable and helps
    \sysname scale to larger distributed topologies. Meanwhile, \sysname still has the freedom to synthesize the
    top-level algorithm and connect the systems to each other as it best can.
\end{example}
\section{Physical Topologies of GPU systems}
\label{sec:topology}

\begin{figure*}[h]
\centering
\begin{subfigure}[t]{0.30\textwidth}
    \includegraphics[width=\textwidth]{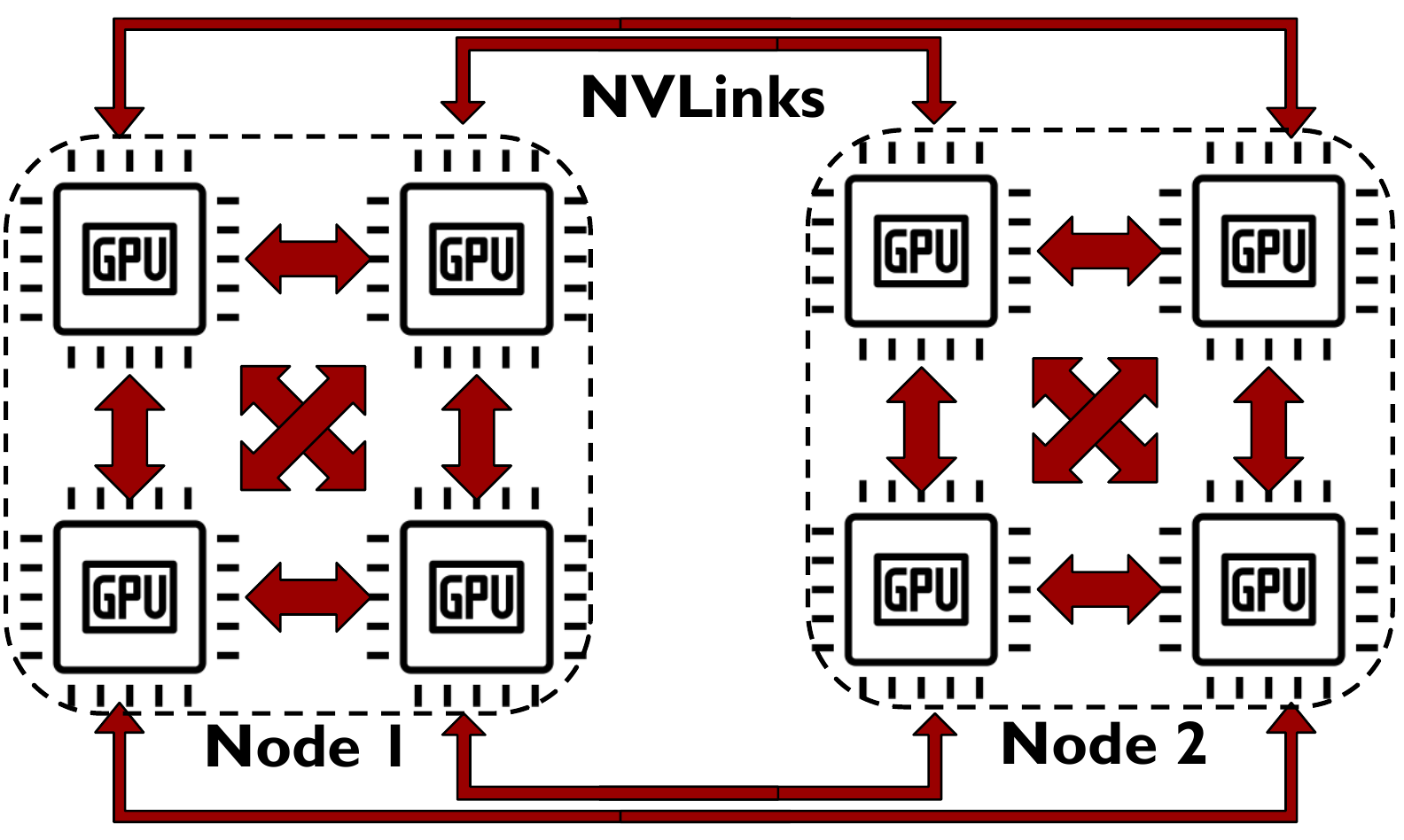}\\
    \caption{NVLink connectivity of a \ndvtwo.}
    \label{fig:ndv2}
\end{subfigure}
\begin{subfigure}[t]{0.35\textwidth}
    \includegraphics[trim=0cm 0cm 0cm 0cm,width=\textwidth]{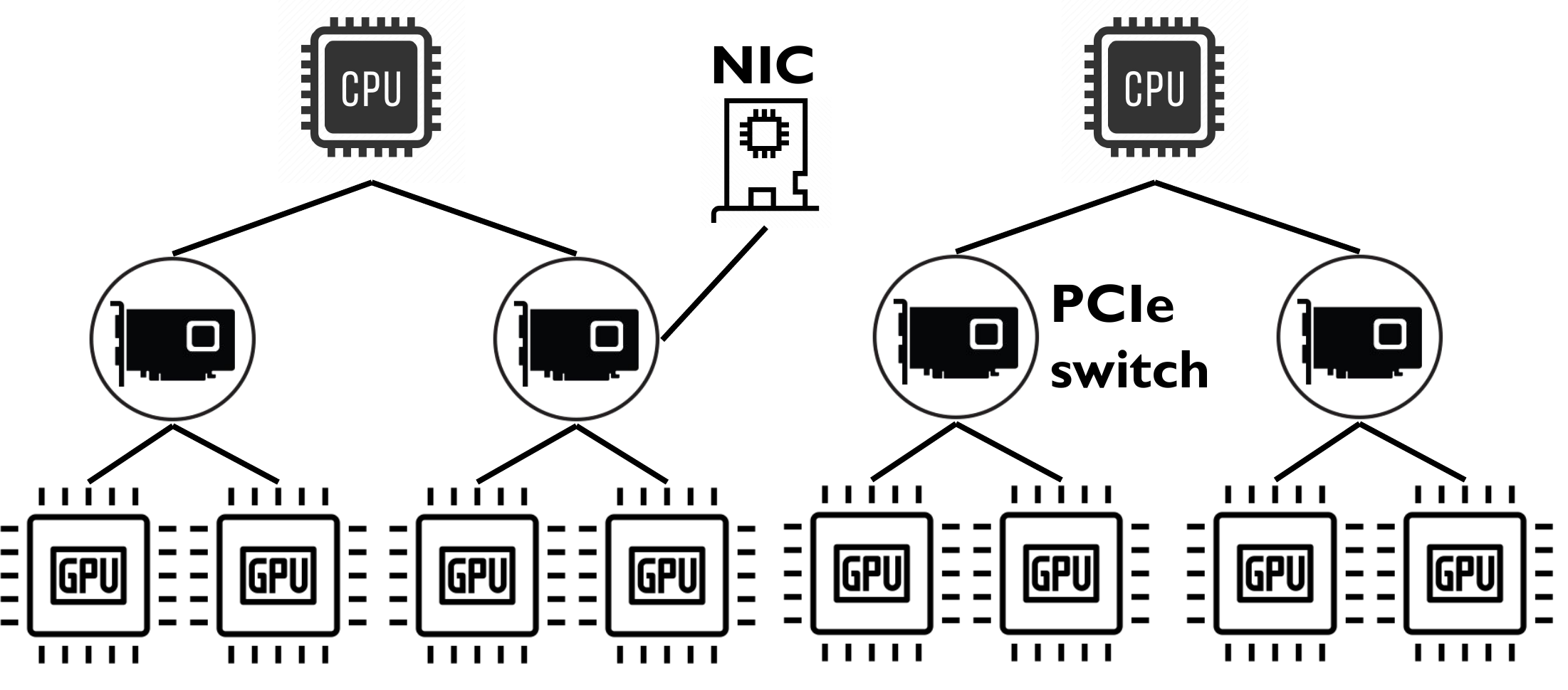}\\
    \caption{PCIe connectivity of a \ndvtwo.}
    \label{fig:ndv2-physical}
\end{subfigure}
\begin{subfigure}[t]{0.32\textwidth}
    \includegraphics[trim=0cm -1cm 0cm 0cm,width=\textwidth]{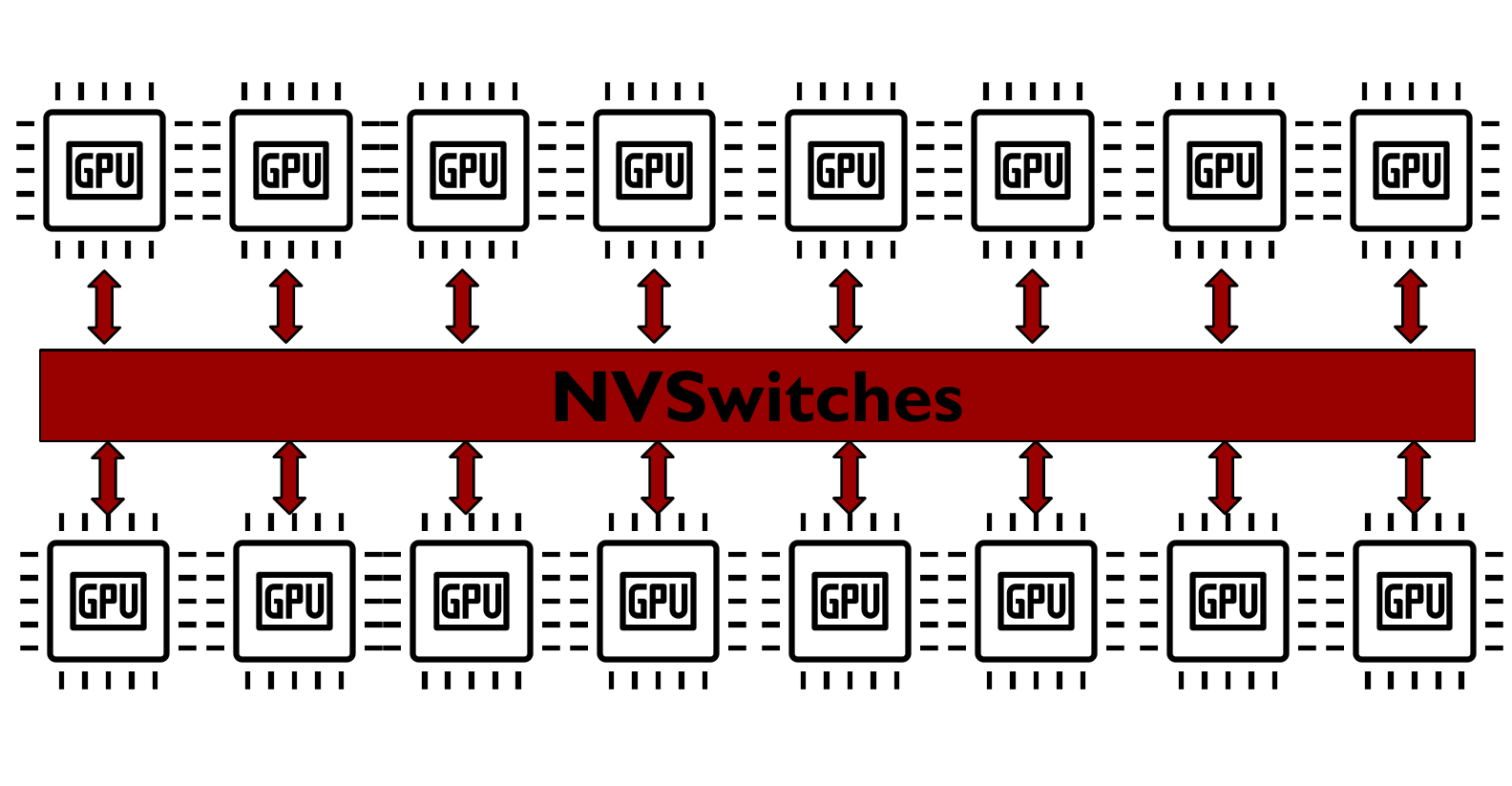}\\
    \caption{NVLink connectivity of a DGX-2.}
    \label{fig:dgx2}
\end{subfigure}
\caption{Aspects of physical topologies in various GPU systems.}
\label{fig:topos}
\end{figure*}

ML engineers use a variety of multi-GPU systems to meet the scaling challenges posed by growing ML
models. Before users can effectively sketch algorithms for \sysname to synthesize, they must understand the
\emph{physical topology} of the target multi-GPU system. However, the performance characteristics of their heterogeneous links
are sparsely documented and for some cloud offerings~\cite{nd-azure} even the topology is not given. To address this,
\sysname includes a {\em physical topology profiler} to measure performance characteristics of links
(\S\ref{subsec:alphabeta}) and to disambiguate the topology of some multi-GPU systems (\S\ref{subsec:topology}).
This section also serves as a concrete introduction into two target systems: Azure \ndvtwo and Nvidia DGX-2.

\subsection{$\alpha$-$\beta$ Cost Model and Link Profiling}
\label{subsec:alphabeta}
In the well-known $\alpha$-$\beta$~\cite{alphabeta} cost model, $\alpha$ is the latency of a link and $\beta$ is the
inverse of its bandwidth. The cost of sending a chunk of size $s$ along a link is $\alpha + \beta \cdot s$.
\sysname's synthesizer adopts the $\alpha$-$\beta$ cost model for simplicity of encoding and tractability, but \sysname's
communication sketches expose features that provide users additional control to avoid excessive concurrency and
congestion (see Section~\ref{sec:sketches}), which are not modeled by the $\alpha$-$\beta$ cost model.
$\alpha$ and $\beta$ are affected by both the interconnect hardware and the software stack running the collective
algorithm (for example software thread fences). \sysname's topology profiler measures and infers the $\alpha$
and $\beta$ costs of different types of links in a GPU system.

Modern GPU systems, e.g., Azure \ndvtwo (Figure~\ref{fig:ndv2}) and
Nvidia DGX-2 (Figure~\ref{fig:dgx2}), have the following types 
of interconnects: (1) \textbf{Peripheral Component Interconnect Express (PCIe)}, 
(2) \textbf{NVLink}~\cite{nvlink}, (3) \textbf{Infiniband} (IB) NICs~\cite{infiniband}. 
A PCIe bus connects GPUs to CPUs with limited shared bandwidth 
(PCIe Gen3 offers $\approx 13$ GBps). PCIe connections often form a 
hierarchy with PCIe switches (Figure~\ref{fig:ndv2-physical}). NVLink~\cite{nvlink}, 
however, is a GPU to GPU \emph{intra-node} connection with dedicated bandwidth.
NVLinks are either directly connected to other GPUs (\ndvtwo in 
Figure~\ref{fig:ndv2}) or they are connected to other GPUs via
NVSwitches~\cite{nvswitch} (DGX2 in Figure~\ref{fig:dgx2}).
NVSwitches enable fully-connected GPU-GPU communication through NVLinks. IB is an 
inter-node interconnect which allows GPUs to communicate with GPUs 
in other nodes like in the Azure \ndvtwo (Figure~\ref{fig:ndv2-physical}). 
IB NICs are usually connected to PCIe switches and GPUs may 
communicate directly with the NICs through Remote Direct Memory Access (RDMA) or
indirectly via host memory.

The profiler empirically derives the $\alpha$ and $\beta$ parameters 
of different links in the network by performing peer-to-peer data transfers 
between GPUs. We send $n$ chunks one after another on a link and measure
the time to transfer. As per the $\alpha-\beta$ cost model, the time to 
transfer is $n \cdot (\alpha + \beta \cdot s)$.  We then send $n$ chunks 
all at once on the link and attribute that time to be $\alpha + n \cdot \beta \cdot s$.
Using several measurements of time to transfer, we solve for $\alpha$ and $\beta$.
Table~\ref{tab:ab-costs} shows the $\alpha$ and $\beta$ values
for \ndvtwo and \dgxtwo systems.
Using these values, we expect that for transfers between two Azure \ndvtwo nodes over InfiniBand (IB),
a sending two 32 KB chunks together as a single 64 KB chunk will be 17\% faster
as compared to sending two 32 KB chunks one after the other.
However, chunks sent together can only be forwarded 
once the last chunk is received. Based on the $\alpha$-$\beta$
values, \sysname's synthesizer determines if and when to 
send chunks together on a link.

\begin{table}[t]
    \small
    \tablestyle{9pt}{1.12}\begin{tabular}{@{\extracolsep{3pt}}lx{20}x{35}x{20}x{35}}
    & \multicolumn{2}{c}{Azure \ndvtwo} & \multicolumn{2}{c}{Nvidia \dgxtwo} \\
    \cline{2-3}  \cline{4-5}
    Link & $\alpha$ (us) & $\beta$ (us/MB) & $\alpha$ (us) & $\beta$ (us/MB)  \\
     \shline
    NVLink  & 0.7   & 46    & 0.7      &   8   \\
    InfiniBand  & 1.7 & 106  & 1.7 & 106  \\
    \end{tabular}
  \caption{Experimentally obtained $\alpha$ and $\beta$ costs for Azure \ndvtwo and Nvidia \dgxtwo nodes.}
    \label{tab:ab-costs}
\end{table}

The $\alpha$-$\beta$ cost model causes \sysname's synthesizer to formulate an 
MILP formulation as opposed to an LP since an algorithm has to be 
expressed in terms of discrete chunks.

\subsection{Inferring Multi-GPU Topologies}
\label{subsec:topology}

For Azure \ndvtwo systems the physical topology was not fully documented: while the NVLink topology
(Figure~\ref{fig:ndv2}) is known to match that of Nvidia DGX1, we did not know how GPUs and the one 12.5 GBps Infiniband
NIC were connected with PCIe. PCIe peer-to-peer communication (and thus GPUDirect RDMA~\cite{gpudirect-rdma}) is not
enabled on these machines, meaning that all communication happens through buffers in CPU memory over potentially shared
PCIe links. Further, virtualization obscures the true PCIe topology (all 8 GPUs and the NIC appear directly connected to
one CPU) and NUMA node and GPU IDs are not assigned consistently from VM to VM. This means that, without additional
information, software cannot avoid contention over shared PCIe links, creating interference and high variance in
performance.

To determine the PCIe topology, \sysname's profiler sends bandwidth and latency probes between the two CPUs, between
pairs of GPUs, and between CPUs and the NIC. It answers the following questions:
\begin{packeditemize}
\item Which CPU is nearest to the NIC? We answer this using the latency 
of loopback operations between the NIC and each CPU.
\item Which GPUs share a PCIe switch? We find all pairs of GPUs that get low bandwidth in a simultaneous copy to the
CPU, indicating contention.
\item Which GPUs share a PCIe switch with the NIC? We find which GPUs get low GPU to CPU bandwidth while the CPU is doing 
a loopback with the NIC. The CPU in this case is the one that is closer to the NIC.
\end{packeditemize}

With this profiling information we were able to deduce the PCIe topology (Figure~\ref{fig:ndv2-physical}). Each CPU has
two PCIe switches connecting to two GPUs each, and the Infiniband NIC is connected to one of these switches.
Additionally, by running the profiler on every new \ndvtwo VM \sysname is able to select one of the NVLink topology's
four automorphisms and set the \texttt{CUDA\_VISIBLE\_DEVICES} environment variable such that the NIC is always placed
close to GPU 0.
\section{\sysname Synthesizer}\label{sec:distsccl-synthesizer}

Once the user has written a communication sketch, they are ready to call \sysname's synthesizer. This section describes
the synthesis process \sysname uses, as well as additional hyperparameters available to the user.

\subsection{Problem Formulation}
GPUs participating in a communication collective partition their initial data into 
$C$ equal chunks where $C$ is a hyperparameter selected by the user. 
\sysname's synthesizer routes and schedules these chunks. Given a communication sketch 
and a collective, the synthesizer decides chunk transfer
schedules across every link in the network, such that each chunk reaches 
its destination GPUs as specified by the collective.

\sysname encodes this problem as a mixed integer linear program (MILP)
with binary and continuous decision variables. The encoding has a continuous variable
called \emph{start\_time} for every chunk and GPU to indicate when a chunk is 
available at a GPU. A binary variable \emph{is\_sent} for all chunk and
link pairs denotes if a chunk is sent over a link. Another continuous variable \emph{send\_time} 
indicates when a chunk is sent over a link. The encoding has bandwidth and 
correctness constraints to ensure the correctness of a chunk transfer schedule. 
The objective of the MILP is to minimize \emph{time} which is a continuous variable 
indicating the maximum time among all chunks that must reach their destination GPUs. 
Details of these variables and constraints are in Appendix~\ref{app:encoding}.

Additionally, \sysname's synthesizer also decides if it should merge some
chunks and transfer them contiguously as one large buffer over a link.
Sending $n$ chunks contiguously in one send instruction over a link requires paying only one 
$\alpha$ latency cost whereas sending $n$ chunks one after the other requires 
paying $n \times \alpha$ latency costs. Note that this does not change the $\beta$ 
bandwidth cost. However, sending $n$ chunks separately over a link enables 
\sysname to order them such that subsequent dependent sends from the 
destination of the link could be scheduled earlier. \sysname's synthesizer 
navigates this trade-off to minimize the time. \sysname uses this feature 
only for IB transfers due to their high $\alpha$ cost and ignores it 
for NVLinks due to their lower latency.

MILP problems in general are NP-hard. Luckily, there are solvers such as Gurobi~\cite{gurobi} 
that apply heuristics to solve MILPs in a feasible way. However, this requires careful
consideration regarding the number of variables and constraints in the formulation.
In \sysname's formulation, transferring chunks over a link cannot overlap and an ordering among them is required.
Therefore, potentially a binary decision is needed for every
pair of chunks that may traverse a link. If we assume there are $C$ chunks for a collective problem,
there are $O(C^2)$ such decisions per link.
Moreover, as the number of nodes increase, the number of links increase linearly (larger topology) 
and the number of chunks for a collective increases linearly (\allgather) or even quadratically (\alltoall). This 
large set of variables and constraints 
leads to infeasible solver time and memory requirements.

To solve this problem, we divide the synthesis into three parts. First, the synthesizer solves an optimization problem
to determine the path used by every chunk without fixing any ordering among chunks, then it heuristically orders the
chunks over every link, and finally, it solves another optimization problem to determine chunk contiguity. 
Complete formal descriptions of each step are in Appendix~\ref{app:encoding}.

\vheading{Step 1: Routing} solves a MILP for finding the path of each chunk independent of other chunks,
allowing chunks sent over a link 
to overlap. The objective of this MILP is to minimize the time, which we constrain to be the maximum of two sets of variables. 
(1) for each link, the number of chunks that traverse that link multiplied by the transfer time of a chunk over that link.
(2) for the path of each chunk, the summation of transfer times of the chunk along every link in the path. Note that
this is only a lower bound on the time since we do not consider link contention or chunk ordering.
\sysname also constrains each chunk's path to be via GPU ranks that are on the shortest paths from their sources
to their destinations using the links the user decided to include in the logical topology. If the communication sketch
specifies an algorithm symmetry, \sysname adds the constraints for the symmetric sends. Replacing switches with
switch-hyperedges is also applied in this step. For each switch-hyperedge, a user-provided policy on the number of unique
connections to/form a switch is applied (see Section~\ref{sec:solver-inclination}).

\sysname uses Gurobi~\cite{gurobi} to solve this MILP and the solution gives every chunk a start\_time for each GPU along its path. 
Clearly this step solves chunk routing, but only partially solves the chunk scheduling and contiguity problem and
requires follow-up steps (explained next) to account for ordering the chunks sent over a link as well as minimizing $\alpha$ costs
of sends.
However, by using this technique, \sysname's synthesizer is able to reduce binary variables
needed from $O(C^2)$ to $O(C)$ per link.

\vheading{Step 2: Heuristic Ordering} decides the chunk ordering sent on each link based on a heuristic. Note that
this step is not an MILP and solely solved by a greedy algorithm. Regardless of when each chunk becomes available at a GPU,
this step assigns a total order on the chunks sent over a link $l=(src,dst)$. This is decided by two heuristic functions.
(1) chunks which need to traverse the longest path from $src$ to their final GPU, have higher priority. (2) 
In case there is tie in (1), chunks which have traversed the shortest path from their initial GPU to $src$,
have higher priority. This ordering will be used in Step 3 to assign the final schedules.

\vheading{Step 3: Contiguity and Exact Scheduling} solves an MILP problem to decide which chunks to send contiguously and gives the exact
schedule. The path to be taken by chunks and their ordering over links have already been determined by the previous
steps which are added as constraints to this MILP. The start\_time and send\_time variables are reassigned in this step by 
considering both the $\alpha$ and $\beta$ costs for each transfer. In this step, the synthesizer allows either sending one
chunk at a time or sending multiple chunks contiguously. This offers a trade-off between (1) sending the chunks that are available at the 
same time for a link according to the ordering in step 2 so that the subsequent sends can be scheduled earlier or (2) sending the chunks
contiguously in one send instruction to save the latency cost. The objective of this MILP is to minimize the total time by 
enforcing all constraints which in \sysname solved by Gurobi~\cite{gurobi}. The solution gives the exact schedule for each chunk.
The details of these constraints and their formulation are in Appendix~\ref{app:encoding}.

\subsection{Synthesizer Hyperparameters}

\sysname's synthesizer has some additional parameters that control the synthesis process. These are
provided by the user to the synthesizer (see Figure~\ref{fig:workflow}) through the communication sketch. 
Details of each parameter is described in Appendix~\ref{app:sketch}.

\myparab{Buffer Size.}
\sysname needs the size of input/output buffers of a collective for the $\alpha$-$\beta$ cost model. 
In ML workloads  the input/output buffer size is a known fixed value.

\myparab{Chunk Partitioning.}
The data buffer at each GPU at the start of the collective can be partitioned into multiple equal chunks.
Each chunk is considered as an atomic scheduling unit by the synthesizer and
different chunks of the same data buffer can be routed over different links.
The semantics of a collective forces a minimum number of chunks such as \alltoall which needs at least
as many chunks as the number of GPU for each buffer. On one hand,
using the minimum number of chunks is often times ideal for finding latency-optimal algorithms.
On the other hand, providing a higher number of chunks 
allows the synthesizer to better utilize the links that might be idle otherwise which is better for finding
bandwidth-optimal algorithms.

\myparab{Switch-Hyperedge Policy.}
\label{sec:solver-inclination}
\sysname can enforce policies for the number of connections established over a set of links in a switch-hyperedge by
counting links utilized for data transfer and setting this count as a part of the MILP objective. The \ucmax policy
will maximize the number of connections, which performs best for small data sizes, while \ucmin will minimize the number
of connections, which works well when the data size is large and congestion is a concern.

\subsection{Synthesizing combining collectives}
\label{sec:combining}
\sysname synthesizes combining collectives (\ie collectives that combine chunks like \reducescatter and \allreduce) by
utilizing synthesis of non-combining collectives, similar to the technique used by SCCL~\cite{cai2021synthesizing}.
\reducescatter can be implemented as an ``inverse'' of \allgather --- a send from a source GPU in \allgather is instead
received and reduced on the source GPU. However, simply inverting the sends does not work --- a GPU may simultaneous
send on different links in an \allgather, but it cannot reduce all receives together in the inverse case. We thus order
the inverse sends using heuristic ordering followed by contiguity encoding in order to synthesize \reducescatter.
\allreduce is synthesized directly by concatenating \reducescatter with an \allgather algorithm.
\section{Backend}\label{sec:backend}

\newcommand{\sysruntime}{\sysname runtime}
\newcommand{\efname}{\sysname-EF}

The synthesizer described above generates an abstract algorithm that specifies the order in which the nodes communicate the various chunks. The goal of the backend is to implement this abstract algorithm. To do so, we extend 
NCCL~\cite{ncclrepo} with an {\em interpreter} which we call \sysruntime{}.
While any communication algorithm can be trivially implemented using 
NCCL's point-to-point sends and receives, \sysruntime{} enables us to execute the entire algorithm in a single kernel launch, eliminating multiple launch overheads. In addition, by reusing NCCL transport mechanisms, \sysruntime\xspace is able to support all of NCCL's communication backends such as IB, Ethernet, NVLink, and PCIe. 

\subsection{\sysruntime{}}

The input to \sysruntime{}\footnote{Link to code: \url{https://github.com/microsoft/msccl}} is a \efname{} program, which is an XML format for representing collective algorithms.
\efname{} programs operate on three buffers: input, output and scratch. 
For each buffer, the program specifies the number of chunks it will be sliced into such that all chunks are equal size. 
Every step of the algorithm is expressed in terms of these chunks.

The program is divided into a set of GPU programs made up of threadblocks.
Each threadblock is made up of a series of steps that are executed sequentially, with each step specifying an instruction and operands as indices into the
input/output/scratch buffers. The current instruction set includes sends, receives (with optional reduction), and local
copies. 
To simplify the implementation of \sysruntime, each threadblock can send to and receive from at most one GPU.  
Additionally, threadblocks within a GPU can synchronize by indicating that one step depends on another step,
which will cause the interpreter to wait until the dependency has completed before executing the dependent step.

The \sysruntime{} extends NCCL and it is backward compatible with its API. Therefore, integrating \sysruntime{} into machine learning 
frameworks such as PyTorch is a single line change wherein that change swaps the third-party NCCL library for \sysruntime{}. This allows \sysname to 
dynamically swap in collective algorithms generated for any training/inference workload using \texttt{torch.distributed}.

\subsection{Lowering to \sysruntime{}}

To target \efname{}, abstract algorithms are lowered to the executable format. The sets of sends operating on
abstract chunks that comprise the steps of the algorithm are transformed into pairs of send and receive
operations operating on concrete buffer indices. Furthermore, these operations are placed sequentially into
threadblocks and any necessary dependencies recorded between them.

\myparab{Buffer allocation.} Input and output buffers are preallocated by the user and passed to the collective. 
Scratch buffers are allocated by the \sysruntime{} per \efname{}. Chunks are indices in the input, output
and scratch buffers. For chunks that are common for both the input and the output buffers (e.g. as in \allgather)
a local copy from input to the output buffer is performed at the end.

\myparab{Instruction generation.} The operations of the abstract algorithm are split into two instructions for the sender
and receiver GPU, and chunks are translated into buffer references and indices according to the buffer
allocation.

\myparab{Dependency insertion.}
\sysname transforms a synthesized algorithm
into the asynchronous execution model of \efname{} and dependencies for each buffer index are inserted 
to ensure that the data dependencies
present in the abstract algorithm are honored.

\myparab{Threadblock allocation.} Instructions are grouped such that all of them are either sending 
to at most one GPU and/or receiving from at most another GPU (possibly different). Order of the instructions inside a group
should follow the order of the abstract algorithm. \sysname allocates a threadblock for each group of instructions.

\myparab{Instances.}
NCCL and consequently \sysruntime{} cannot saturate the bandwidth of a link in a topology using a single threadblock.
Thus, \sysname{} generates multiple instances of the algorithm to maximize the performance. This
is done by subdividing each chunk into $n$ subchunks that follow the same path as the parent chunk. 
All groups of instructions and their threadblocks are duplicated $n$ times and executed in parallel.
\S\ref{sec:eval-impact} explores the performance implications of choices of $n$.

\section{Evaluation}\label{sec:eval}

We evaluate algorithms obtained with \sysname
for \allgather, \alltoall, and \allreduce collectives on a cluster
of 32 GPUs comprised of two Nvidia \dgxtwo nodes or up-to four Azure \ndvtwo nodes.
To compare performances, algorithm bandwidth~\cite{nccltests} measurement is used which
is calculated by input buffer size divided by execution time.
We synthesize \sysname algorithms by exploring different communication sketches
and compare them
against the popular Nvidia Collective Communication Library (NCCL) (v.2.8.4-1).
This section analyzes how different communication sketches impact the performance of the algorithms
synthesized by \sysname. In particular, we perform ablation studies by varying the 
inter-node connections in the logical topology, changing synthesizer hyperparameters,
and changing the number of instances used when lowering to \sysname-EF.
To evaluate how \sysname's speedups translate to end-to-end performance,
we use algorithms generated by \sysname in two large language models, Transformer-XL and BERT.
Finally, we discuss the synthesis time required by \sysname to generate these algorithms.

We believe our focus on up to 32 GPUs covers a large section of important use cases: in an internal cluster of \dgxtwo
nodes at Microsoft, the sum of GPUs in jobs of at most 32 was 93.7\% of all jobs in the second half of 2021.
  
\subsection{Standalone Experiments}
\label{sec:standalone}

\begin{figure}[!htb]
  \centering
  \includegraphics[width=\linewidth]{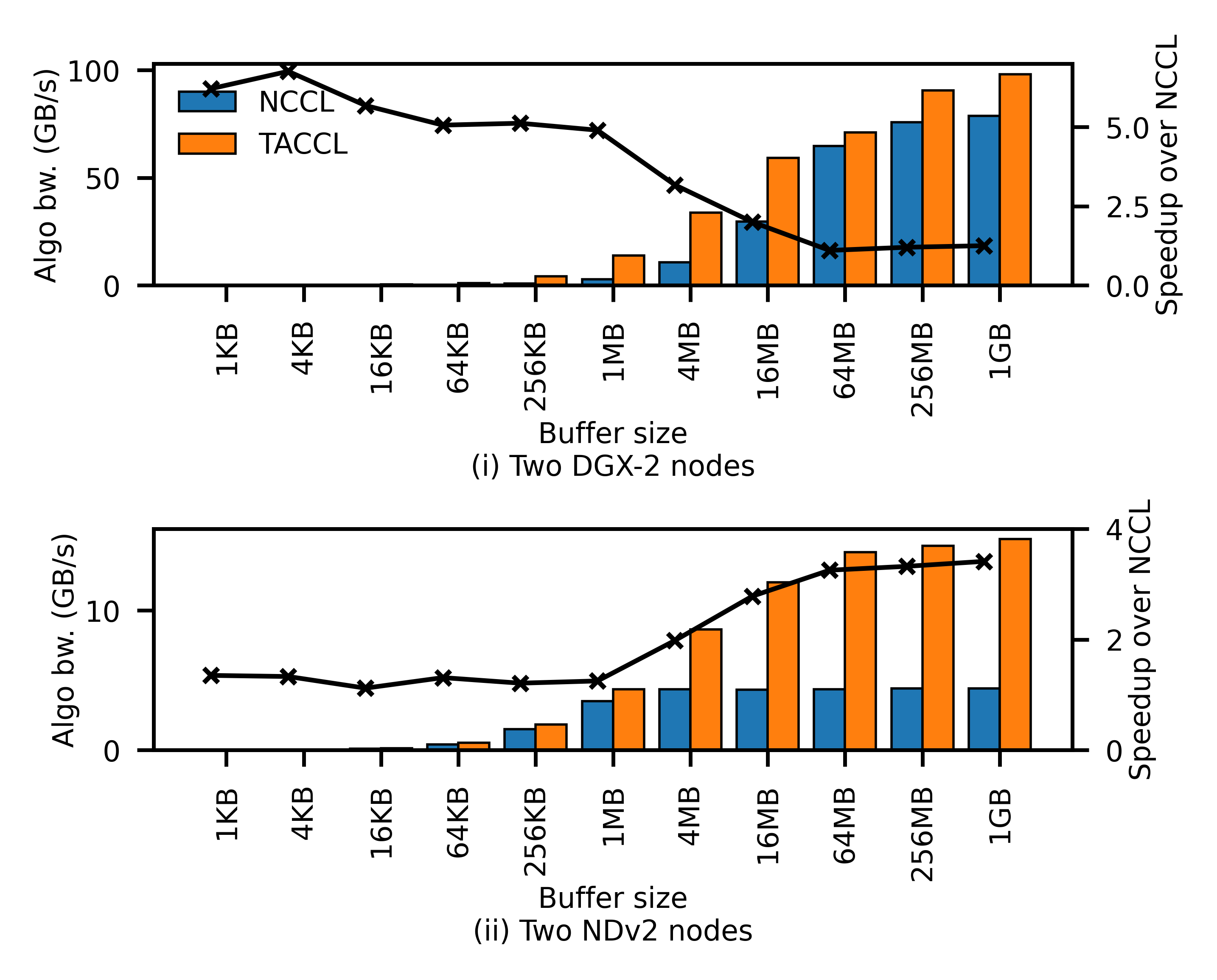}
  \caption{\allgather comparisons of NCCL to \sysname's best algorithm at each buffer size.}
  \label{fig:exp-allgather}
\end{figure}

All our communication sketches for \dgxtwo and \ndvtwo use a hierarchical symmetry like the one in
Example~\ref{ex:ndv2-symmetry}.

\subsubsection{\allgather}
\myparab{\allgather on \dgxtwo.}
Figure~\ref{fig:exp-allgather}(i) shows the algorithm bandwidth for \sysname's synthesized algorithms on two \dgxtwo nodes for each output buffer size
and plots it against that of NCCL.
We show the speedup of \sysname's algorithms over NCCL on the right Y-axis of the plot. We used two different sketches 
for this topology which will be explained next.

A \dgxtwo node has 16 V100 GPUs (Figure~\ref{fig:dgx2}) where each pair of GPUs share a PCIe switch with a NIC. This
makes it natural to assign
one GPU in a pair to be a receiver and the other to be a sender by eliminating outgoing and
incoming links, respectively, in the logical topology.
We design a sketch (\emph{dgx2-sk-1}) that uses this logical
topology, sets chunk size to $2$MB, 
uses two chunk partitions for each buffer, and the 
sets switch-hyperedge policy to \ucmin. With this sketch,
\sysname synthesizes an \allgather algorithm for two \dgxtwo nodes.
This algorithm almost saturates the inter-node bandwidth during the entire run of the algorithm
and provides a $20\% - 25\%$ speedup
over NCCL for large buffer sizes in the $256$MB - $1$GB range.

Next, we design a sketch (\emph{dgx2-sk-2}) for smaller sizes. This sketch allows both GPUs in a pair 
to utilize the shared NIC. However, local GPU $i$ on each node is only allowed to send/receive to/from local GPU $i$ on the other node. 
Since the IB is shared, we double the $\beta$ cost for each IB transfer to $2*\beta_{IB}$ cost.
In this sketch, chunk size is set to $1$KB and the switch-hyperedge policy is \ucmax.
Using this sketch \sysname synthesizes an algorithm that is $4.9\times - 6.7\times$ faster than NCCL in the $1$KB - $1$MB range,
and $10\% - 3.8\times$ faster than NCCL in the $2$MB - $64$MB range.
On inspecting this algorithm, we found that \sysname's synthesized algorithm 
overlaps inter-node sends with intra-node all-pair \allgather of node-local data chunks followed by
an intra-node all-pair \allgather of the node-external chunks received over IB.

Figure~\ref{fig:exp-allgather}(i) shows the algorithm bandwidth and the speedup over NCCL baseline for the best of these two sketches for each output buffer size.

\myparab{\allgather on \ndvtwo.}
The sketch we used, \emph{ndv2-sk-1}, uses the logical topology discussed in Example~\ref{ex:dist-ndvtwo}, in which
a sender and a receiver GPU were dedicated such that they are on the same PCIe switch as the NIC.
We use a single instance when lowering algorithms into \efname{}
for data sizes $1$MB and below, and use 8 instances for larger data sizes.
Figure~\ref{fig:exp-allgather}(ii) compares the synthesized algorithms to NCCL on two Azure \ndvtwo nodes.
\sysname's synthesized algorithms are  $12\% - 35\%$ faster than NCCL for buffer sizes of $1$KB - $1$MB,
and $61\% - 3.4\times$ faster than NCCL for sizes larger than $1$MB.
These algorithms 
better saturate the inter-node bandwidth thanks to the dedicated send/receiver GPUs.

We similarly synthesize \allgather algorithms for four \ndvtwo nodes and present the results in
Figure~\ref{fig:ndv2-app}(i) in Appendix~\ref{app:four-node}.
These algorithms are $10\% - 2.2\times$ faster than NCCL depending on buffer size.

\begin{figure}[!tb]
  \centering
      \includegraphics[width=\linewidth]{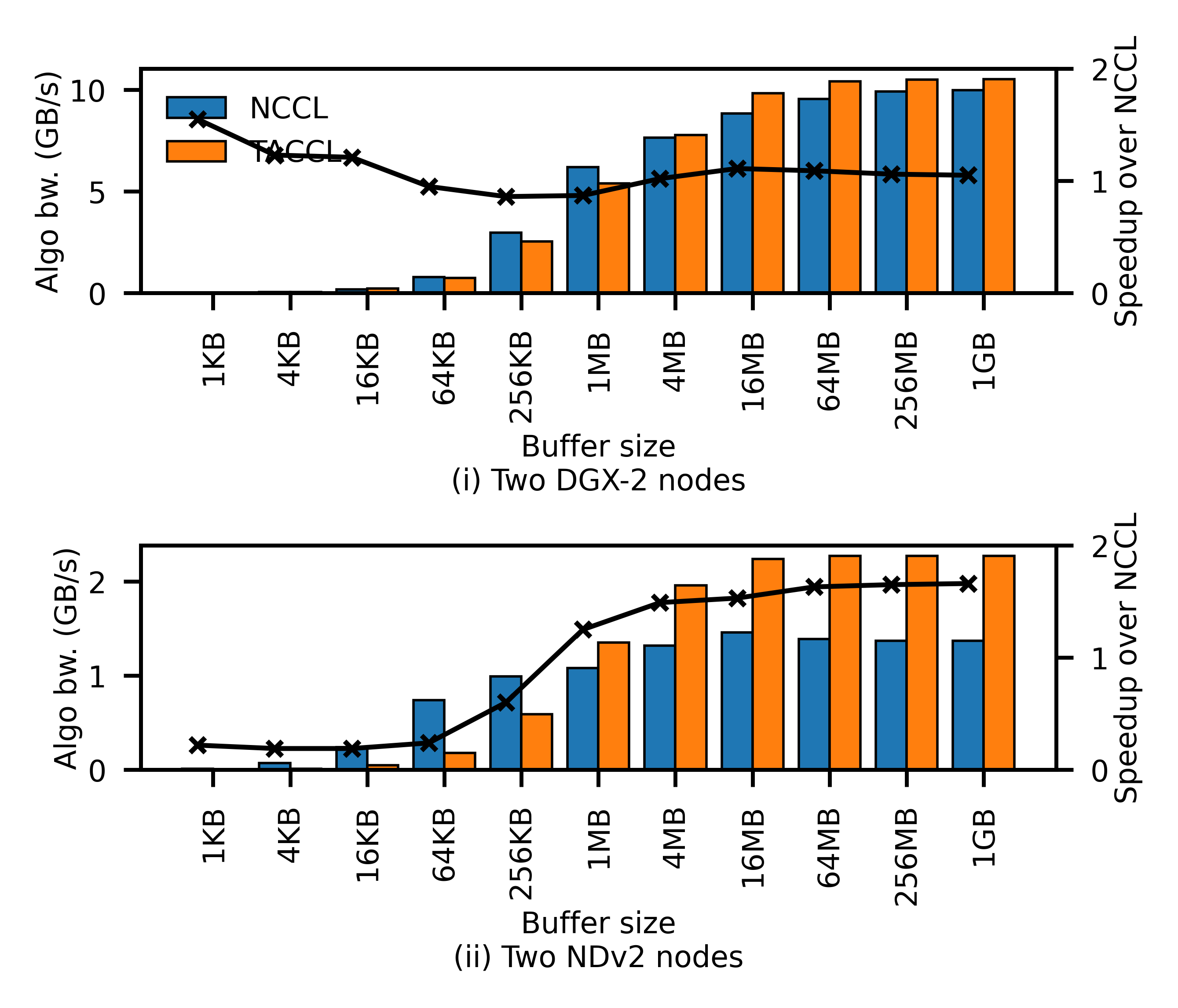}
    \caption{\alltoall comparisons of NCCL to \sysname's best algorithm at each buffer size.}
    \label{fig:exp-alltoall}
  \end{figure}

\subsubsection{\alltoall}
\myparab{\alltoall on \dgxtwo.}
We explore the synthesis of \alltoall algorithms by reusing the \emph{dgx2-sk-2} communication sketch
designed in the previous section.
Figure~\ref{fig:exp-alltoall}(i) compares the resulting algorithm on two \dgxtwo nodes.
The synthesized algorithm using this sketch performs up-to $15\%$ faster than NCCL for
batch sizes of $2$MB and larger. For this sketch,
\sysname's synthesizer coalesces chunks sent in inter-node transfer in this algorithm, which reduces the latency
of transfers over IB.
\sysname also uses a communication sketch with chunk size set as $1$KB and a logical topology 
where 
GPUs have links to all other GPUs connected via the NIC (\emph{dgx2-sk-3}).
This algorithm is up-to $55\%$ faster than NCCL for small buffer sizes ranging from $1$KB to $16$KB.

\myparab{\alltoall on \ndvtwo.}
Figure~\ref{fig:exp-alltoall}(ii) shows a comparison of \sysname's best algorithms for \alltoall on
two Azure \ndvtwo nodes against NCCL.
We reuse the communication sketch \emph{ndv2-sk-1} and set the chunk size to $1$MB.
The generated algorithms  run $53\% - 66\%$ faster than NCCL for buffer sizes between $16$MB - $1GB$
We explore another sketch (\emph{ndv2-sk-2}) with a logical topology in which all GPUs in a node are fully-connected
to all the GPUs in the other node and set chunk size as $1$KB.
The algorithm generated by \sysname using this sketch performs up-to $12\%$ faster than
NCCL for buffer sizes from $1$KB to $128$KB.

For four \ndvtwo nodes, \sysname's synthesized algorithms  uses communication sketch \emph{ndv2-sk-1} and they are
up-to $46\%$ faster than NCCL for buffer size greater than $1$MB, as shown in Figure~\ref{fig:ndv2-app}(ii) in Appendix~\ref{app:four-node}.

\begin{figure}[!tb]
  \centering
      \includegraphics[width=\linewidth]{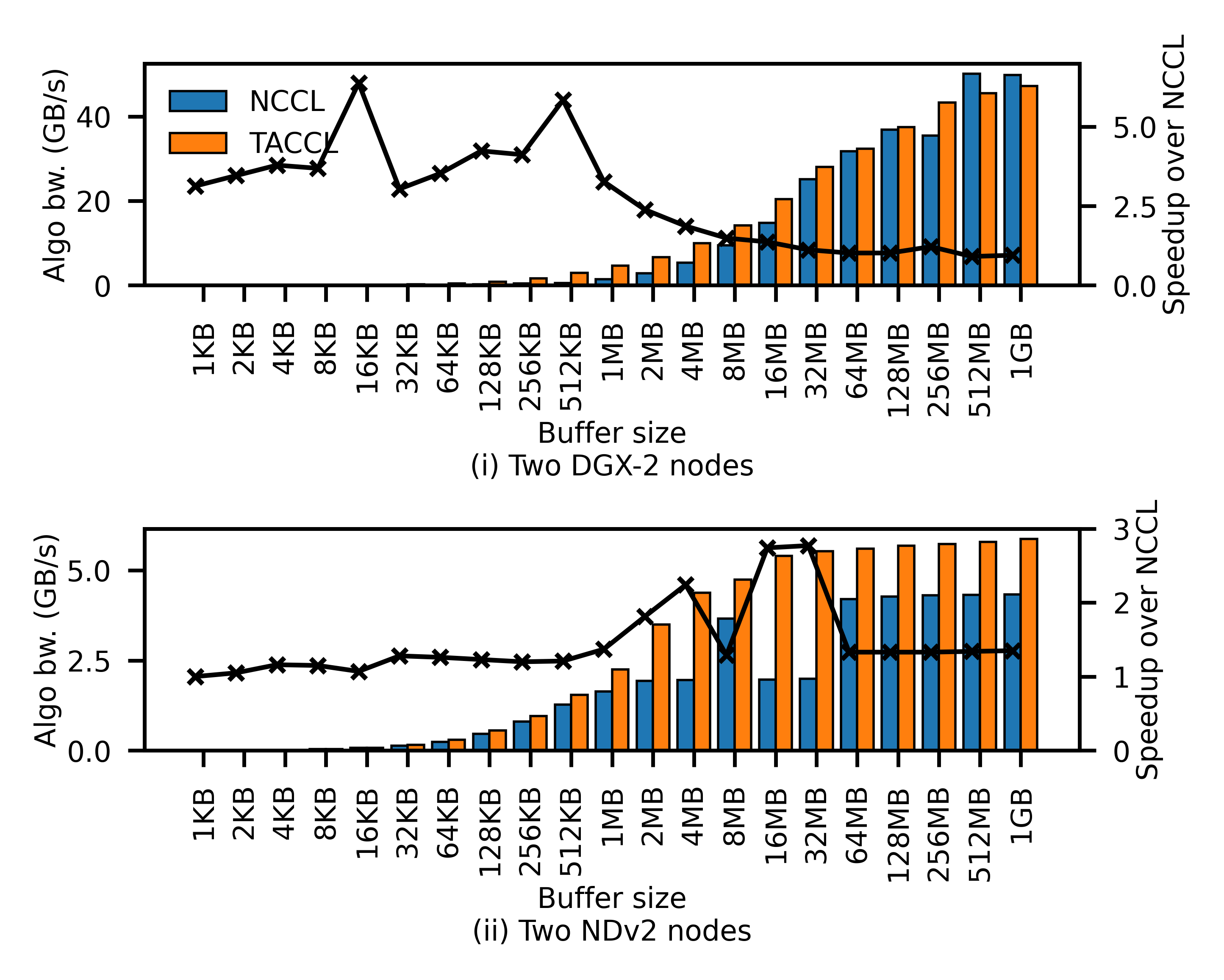}
    \caption{\allreduce comparisons of NCCL to \sysname's best algorithm at each buffer size.}
    \label{fig:exp-allreduce}
  \end{figure}

\subsubsection{\allreduce}
\myparab{\allreduce on \dgxtwo.}
As discussed in Section~\ref{sec:combining}, \sysname composes \reducescatter with \allgather to implement \allreduce
and an algorithm for \reducescatter can be constructed by inverting an \allgather algorithm.
Figure~\ref{fig:exp-allreduce}(i) shows the performance of \sysname algorithms on two \dgxtwo nodes.
The \allreduce synthesized from the \allgather using \emph{dgx2-sk-2} is $49\% - 6.4\times$ faster than NCCL for
buffer sizes ranging from $1$KB - $4$MB.
\sysname's generated algorithms by using other communication sketches like \emph{dgx2-sk-1} are $2\% - 37\%$ faster
than NCCL for buffer sizes ranging from $16$MB - $256$MB.
For buffer sizes of $512$MB and greater, our algorithms are at most $9\%$ slower than NCCL.
This is because NCCL uses the more optimized fused communication instructions (such as receive-reduce-copy-send) in its \allreduce
communication which are unavailable in \sysname's lowering. We leave these such further optimizations for future work.

\myparab{\allreduce on \ndvtwo.}
These algorithms are based on the \allgather synthesized from the \emph{ndv2-sk-1} sketch and
use two versions with 1 and 8 instances.
Figure~\ref{fig:exp-allreduce}(ii) compares them to NCCL on two \ndvtwo nodes. 
The single instance \sysname algorithm outperforms NCCL's \allreduce by up to $28\%$ for buffer sizes of up to $1$MB,
while the 8 instance algorithm outperforms NCCL by $28\% - 2.7\times$ for larger sizes.

On 4 \ndvtwo nodes, as shown in Figure~\ref{fig:ndv2-app}(iii) in Appendix~\ref{app:four-node}, the \sysname algorithms are up to $34\%$ faster than NCCL
for small buffer sizes and $1.9\times - 2.1\times$ faster than NCCL for larger buffer sizes.

\begin{figure*}[!htb]
  \centering
  \begin{subfigure}[h]{0.44\textwidth}
    \includegraphics[width=0.97\textwidth]{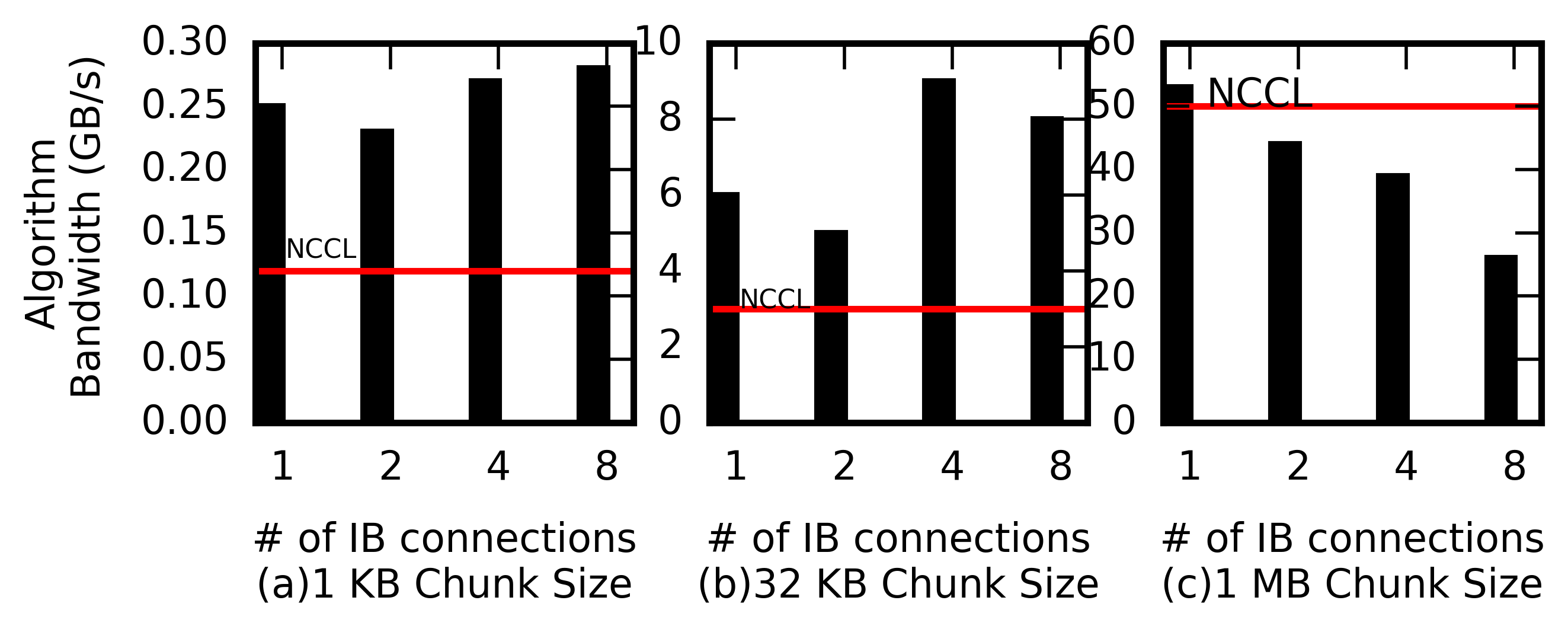}
    \caption{Logical topology}
  \label{fig:ib-ablation}
\end{subfigure}%
  \begin{subfigure}[h]{0.52\textwidth}
    \includegraphics[width=0.97\textwidth]{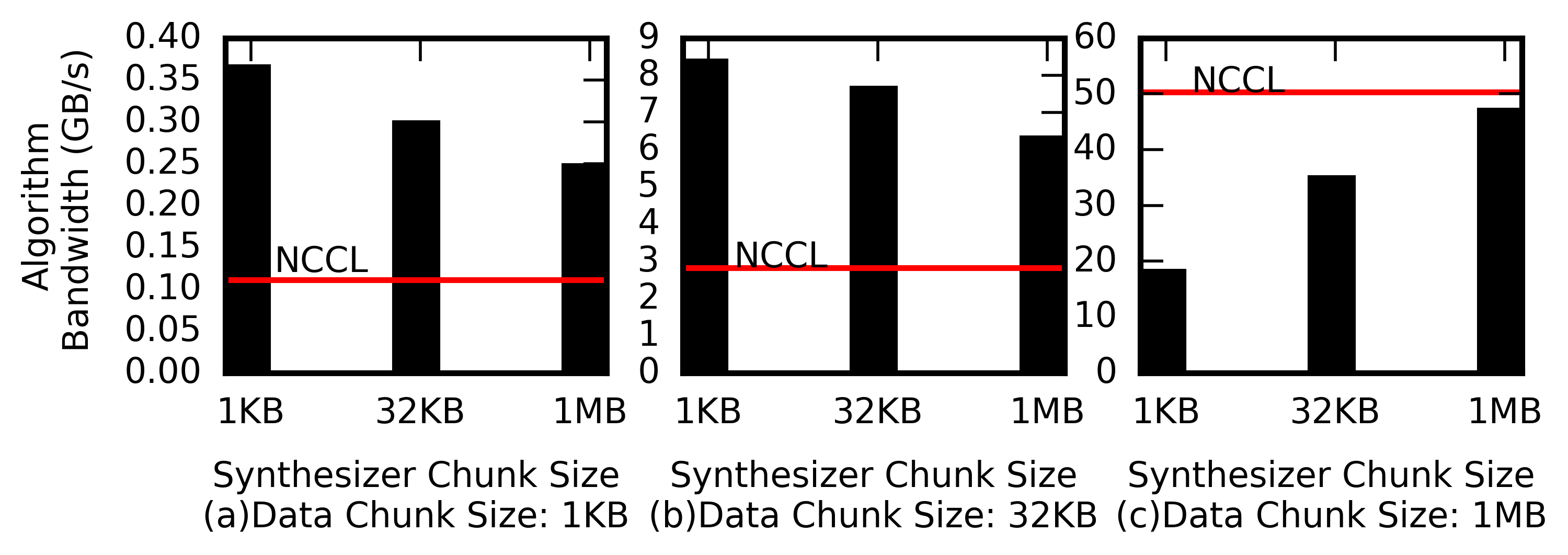}
    \caption{Chunk size}
  \label{fig:chunksz-ablation}
\end{subfigure}%

\begin{subfigure}[h]{0.20\textwidth}
  \includegraphics[height=1.3in]{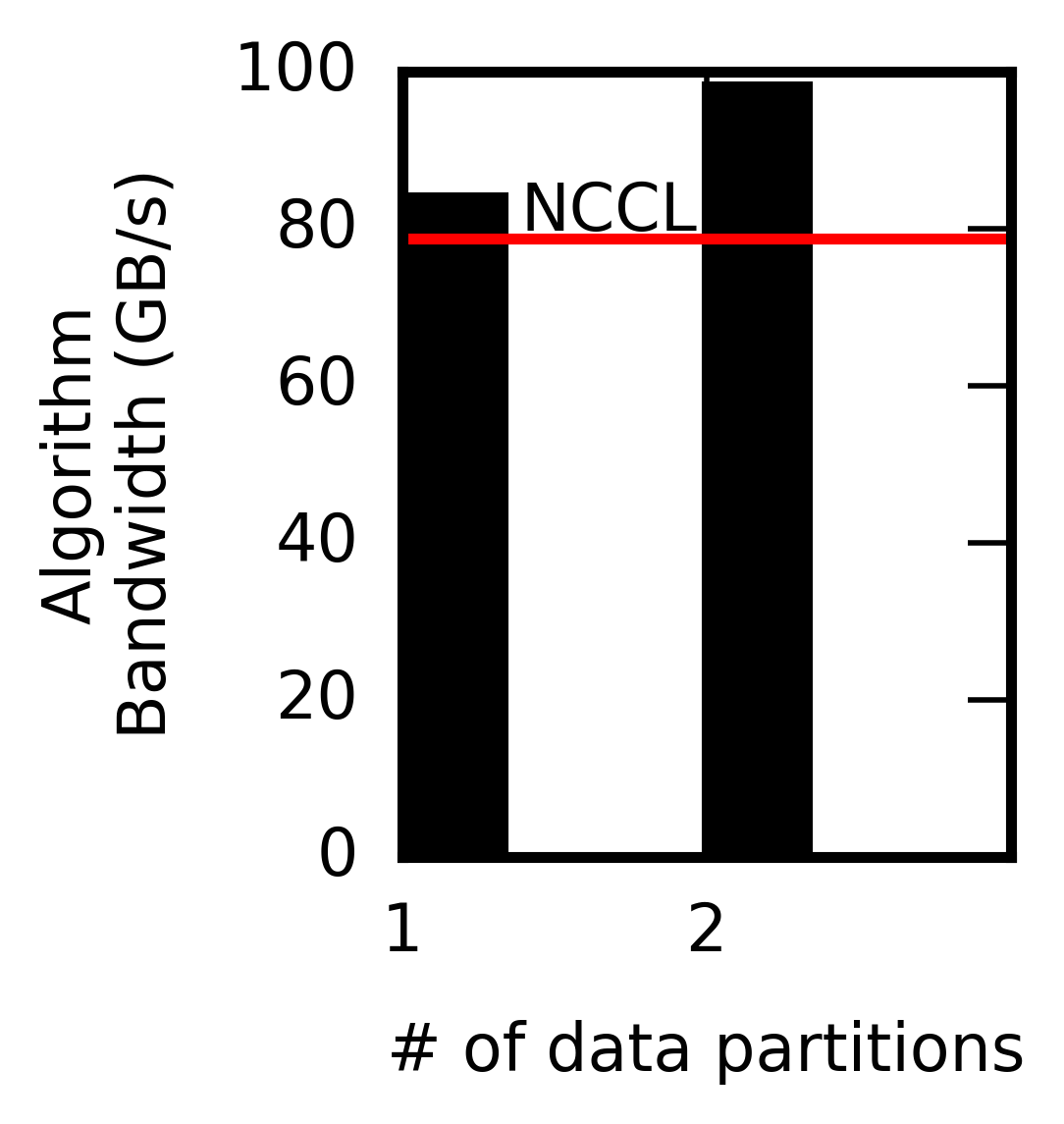}
  \caption{Data partition}
\label{fig:chunkup-ablation}
\end{subfigure}%
  \begin{subfigure}[h]{0.37\textwidth}
    \includegraphics[height=1.3in]{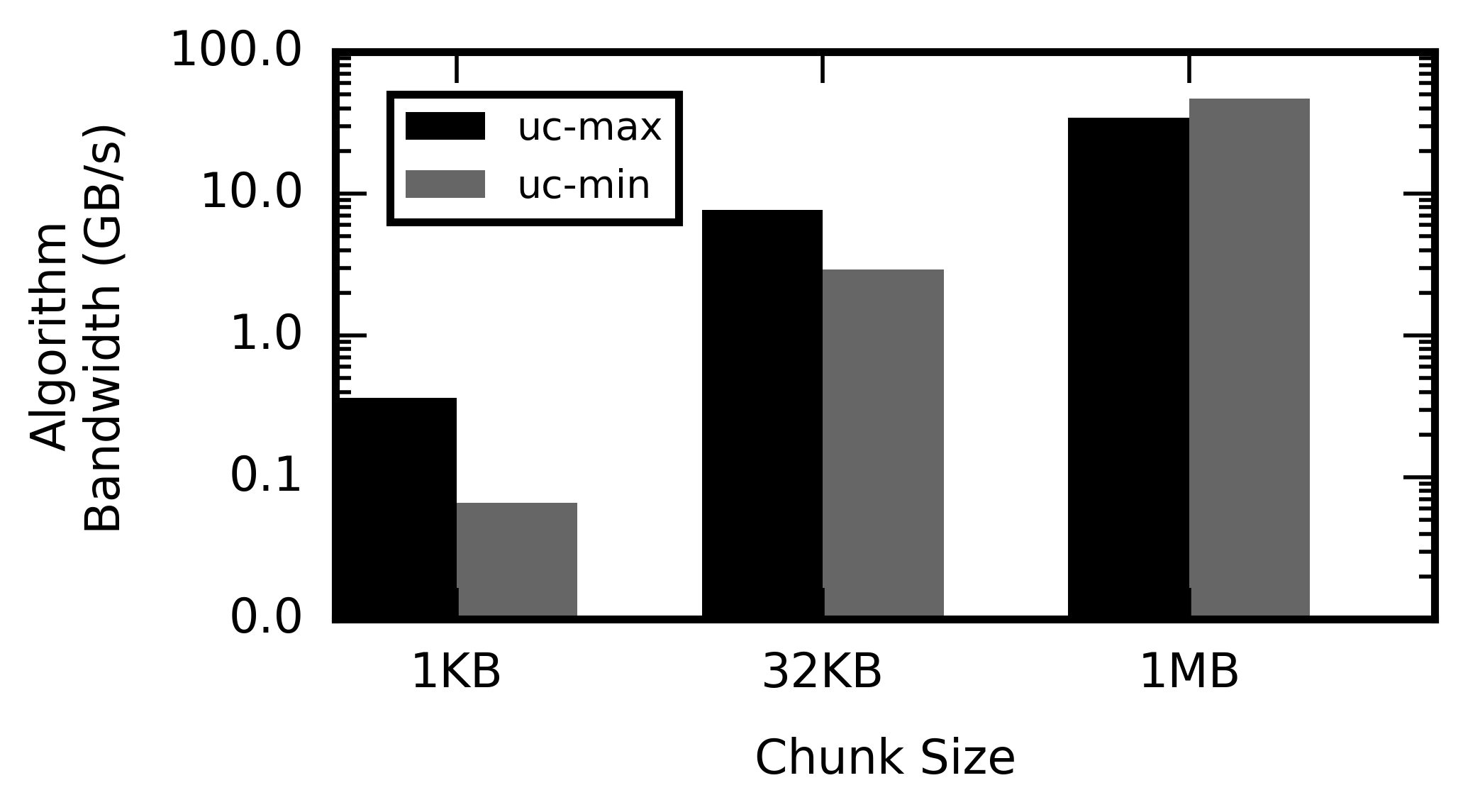}
    \caption{Switch-hyperedge strategies}
  \label{fig:uc-ablation}
\end{subfigure}%
  \begin{subfigure}[h]{0.37\textwidth}
    \includegraphics[height=1.3in]{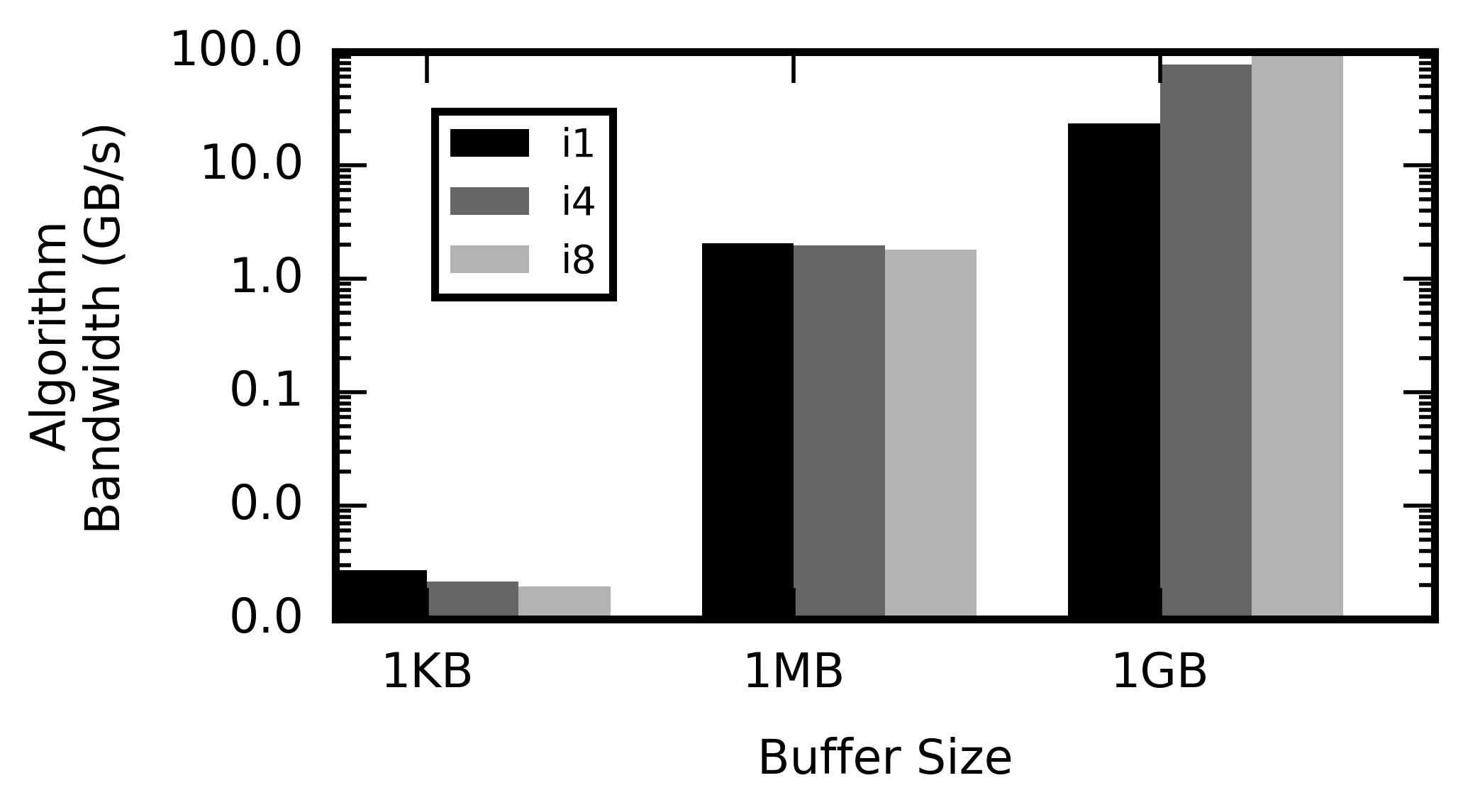}
    \caption{Runtime instances}
  \label{fig:i-ablation}
\end{subfigure}%
\caption{\small{Algorithm bandwidth of \allgather algorithms on \dgxtwo by varying different inputs to \sysname}}
  \label{fig:all-ablation}
\end{figure*}

\subsection{Impact of Varying Synthesizer Inputs}\label{sec:eval-impact}
In this section, we explore modifications to communication sketches,
as well as the synthesizer hyperparameters and the instances for the lowering,
in order to understand their impact on the performance of the synthesized algorithms.
Our aim is to demonstrate that the controls offered by \sysname have intuitive effects
on the resulting algorithms, which is necessary for effectively communicating
user intuition to \sysname.

We present our analysis for the \allgather collective on two Nvidia \dgxtwo nodes.
Unless mentioned otherwise, we use the following communication sketch as the baseline: same logical topology
as \emph{dgx2-sk-1}, chunk size set to $1$MB, data partitioning set to 1, and the switch-hyperedge policy set to \ucmax.

\myparab{Changing logical topology.}
We create a logical topology with a dedicated sender and receiver GPU similar to \emph{dgx-sk-1}
except we allow a sender to be connected to $n$ different receivers in the other node.
Figure~\ref{fig:ib-ablation} shows the algorithm bandwidth of \allgather obtained by varying $n$,
the number of IB connections per GPU, for a fixed chunk size of $1$KB, $32$KB, and $1$MB.
For a $1$KB chunk size,
we found the algorithm that uses 8 IB connections per NIC performs better than algorithms using fewer connections.
As the chunk size increases to $32$KB and $1$MB, the optimal number of
IB connections per NIC reduces to 4 and 1, respectively.
The benefits of link sharing shrink as the chunk size increases and $\beta$-cost starts
dominating over the $\alpha$-cost.

\myparab{Changing transfer cost using chunk size.}
We analyze the sensitivity of \sysname's synthesizer to the data size provided in the communication
sketch when its algorithms are applied on a communication using a different data size.
Figure~\ref{fig:chunksz-ablation} shows the performance of \allgather algorithm for three
different chunk sizes ($1$KB, $32$KB, and $1$MB).
Algorithms generally perform well for a range of data sizes close to what they have been synthesized for.
We recommend trying a small set of nearby sizes to ensure the best performance.

\myparab{Changing data partitioning.}
Figure~\ref{fig:chunkup-ablation} shows the algorithm bandwidth of algorithms generated by
partitioning data on each GPU into a single or two chunks.
We set the switch-hyperedge policy to \ucmin and fix number of instances to 8.
At a large buffer size of $1$GB, the algorithm generated for two data chunks utilizes bandwidth
better as compared to the algorithm generated for a single data chunk per GPU.

\myparab{Changing switch-hyperedge policy.}
Figure~\ref{fig:uc-ablation} shows the algorithm bandwidth for algorithms generated
and evaluated for $1$KB, $32$KB, and $1$MB chunks.
The algorithm bandwidth is displayed in log-scale.
We vary the switch-hyperedge policy between \ucmax and \ucmin.
For smaller buffer sizes, the \ucmax configuration performs better than \ucmin,
whereas for larger buffer sizes, \ucmin performs better than \ucmax.

\myparab{Changing number of instances.}
Figure~\ref{fig:i-ablation} shows algorithm bandwidth with instances ranging from 1 to 8.
The switch-hyperedge policy for these algorithms is set to \ucmin.
Increasing the number of instances improves bandwidth utilization \textemdash{} multiple
threadblocks seem to be needed to keep the six NVLinks in a V100 busy.
However, a larger number of threadblocks also increases latency, which we suspect is due to
unfavorable scheduling of synchronization related memory operations onto the NVLinks at the start of each send.
Since latency cost dominates for small buffer sizes, using a large number of instances only increases the latency cost.
As the buffer size increases, the bandwidth improvements due to more instances become predominant.
Since switch-hyperedge policy and number of instances have a similar relation with chunk sizes, we always run
\ucmax algorithms with a single instance and \ucmin algorithms with 8 instances.

\begin{figure}[!htb]
  \centering
  \begin{subfigure}[h]{0.9\linewidth}
    \includegraphics[width=0.97\linewidth]{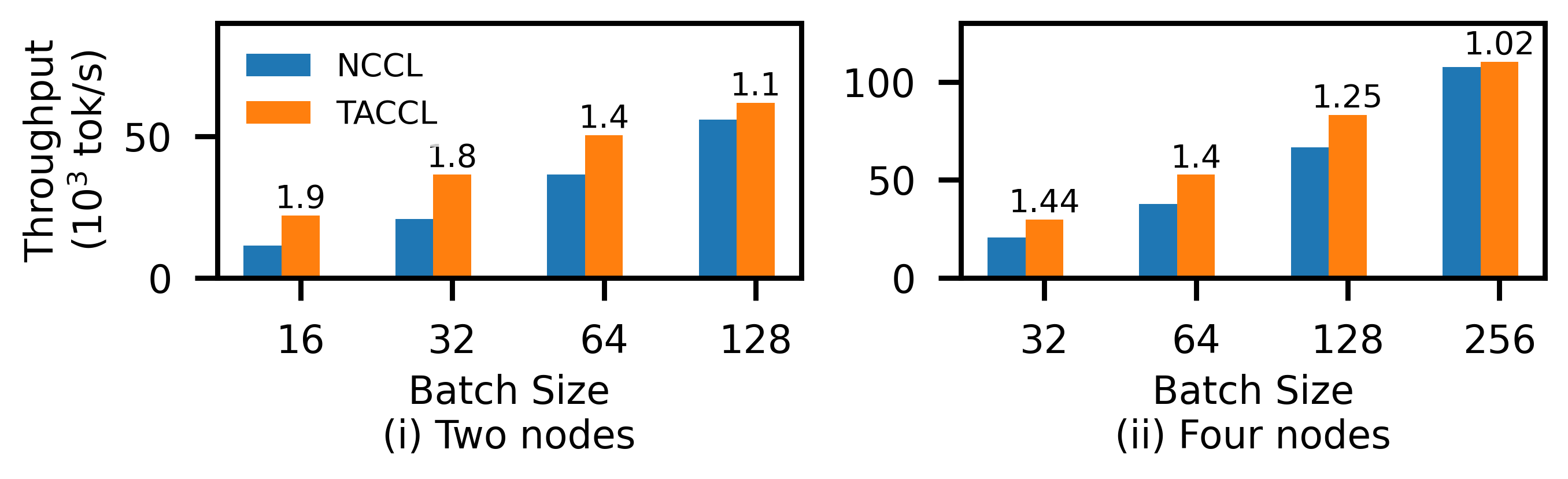}
    \caption{Transformer-XL}
  \label{fig:transformer}
\end{subfigure}%

  \begin{subfigure}[h]{0.9\linewidth}
    \includegraphics[width=0.97\linewidth]{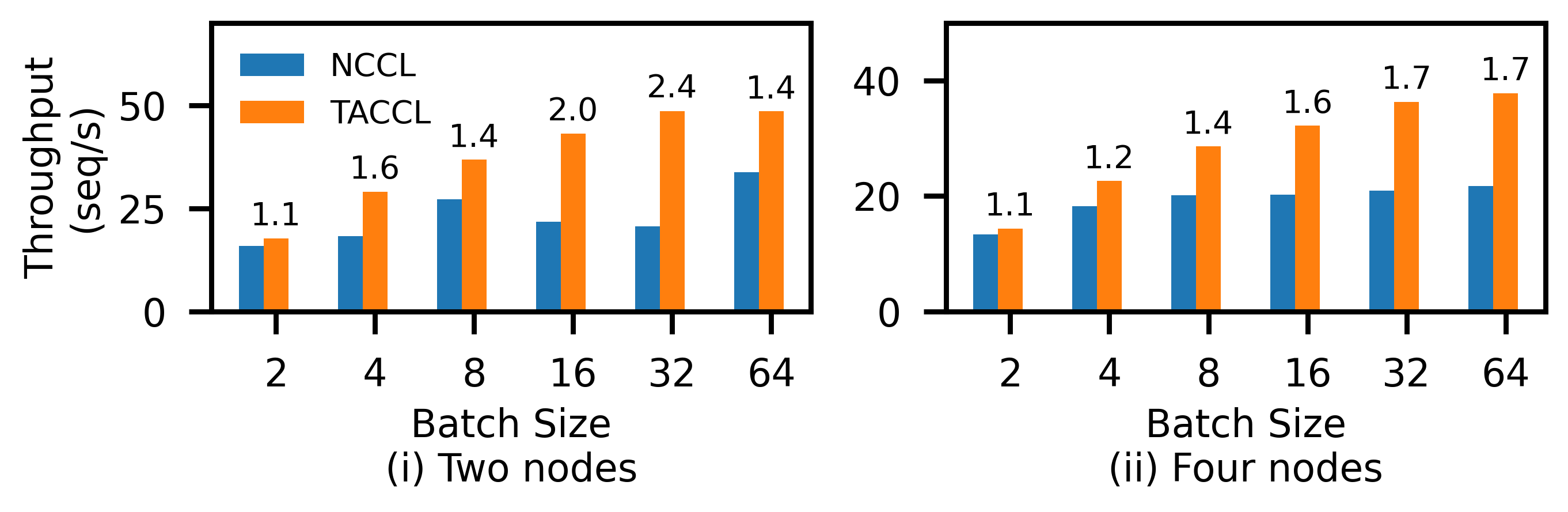}
    \caption{BERT}
  \label{fig:bert}
\end{subfigure}%
\caption{Training throughput using \sysname's collective algorithms on Transformer-XL and BERT
compared against NCCL
on 2 and 4 Azure \ndvtwo nodes. Speedup over NCCL is mentioned
on top of the bars.}
\label{fig:e2e}
\end{figure}

\subsection{End-to-End Training.}
We evaluate \sysname on distributed training of two large language models,
Transformer-XL~\cite{dai2019transformer,transformerCode} and BERT~\cite{bert,bertCode},
on two (and four) Azure \ndvtwo nodes, i.e. 16 (and 32) GPUs.
Transformer-XL uses data parallelism and
whereas BERT uses model parallelism.
The typical transfer sizes for \allreduce in Transformer-XL is in the
$20$ - $40$MB range, and for BERT it is about $2$MB.
Both models communicate with 
\texttt{torch.distributed} and, as explained in Section~\ref{sec:backend},
using \sysname algorithms in them is quite straightforward.

We lower the algorithm synthesized by the synthesizer into \sysname-EF with
1 and 8 instances, and show the performance of both against NCCL.
Figure~\ref{fig:transformer} and Figure~\ref{fig:bert} show the training throughput
obtained by using \sysname's collective algorithms for communication instead of NCCL
for Transformer-XL and BERT respectively for different batch sizes.
\sysname speeds up training of Transformer-XL by $11\% - 1.94\times$ on 2 nodes
and by $2\% - 1.44\times$ on 4 nodes. The speedup for BERT is $12\% - 2.36\times$ on 2 nodes
and $7\% - 1.74\times$ on 4 nodes.
Depending on the memory available per GPU and on how the batch size affects model accuracy,
any of these batch sizes might be chosen for use in practice.

We also use algorithms synthesized by \sysname for \alltoall and \allreduce collectives
for training an internal Microsoft's mixture-of-experts workload on two \ndvtwo nodes.
The \alltoall and \allreduce sizes required for this model are $\approx 6$MB and 
$\approx 256$MB, respectively. \sysname improves the end-to-end throughput of this model by 17\%.

\begin{table}[t]
  \footnotesize
  \centering
  \begin{tabular}{ ccc }
    \begin{tabular}{cM{0.5cm}M{0.6cm}c} \hline
    \multicolumn{2}{c}{AllGather} \\
    \hline
    Sketch & Time(s) \\
    \hline
    \emph{dgx2-sk-1}    & 35.8               \\
    \emph{dgx2-sk-2}    & 11.3               \\
    \emph{ndv2-sk-1}   & 2.6                \\
    \end{tabular}
    \begin{tabular}{cM{0.5cm}M{0.6cm}c} \hline
    \multicolumn{2}{c}{AlltoAll} \\
    \hline
      Sketch & Time(s) \\
      \hline
    \emph{dgx2-sk-2}      & 92.5               \\
    \emph{ndv2-sk-1}     & 1809.8              \\
    \emph{ndv2-sk-2}     & 8.4                \\
    \end{tabular}
    \begin{tabular}{cM{0.5cm}M{0.6cm}c} \hline
    \multicolumn{2}{c}{AllReduce} \\
    \hline
      Sketch & Time(s) \\
      \hline
    \emph{dgx2-sk-1}      & 6.1                \\
    \emph{dgx2-sk-2}      & 127.8              \\
    \emph{ndv2-sk-1}     & 0.3                \\
    \end{tabular}
  \end{tabular}
  \caption{Synthesis time for \sysname algorithms for
  different collectives using different communication sketches.}
  \label{tab:synth-time}
\end{table}

\subsection{Synthesis Time}
Table~\ref{tab:synth-time} shows the total time it takes for \sysname to synthesize algorithms
for different collectives using some of the communication sketches mentioned in
Section~\ref{sec:standalone}.
In most cases synthesis takes from seconds to a few minutes, making it amenable to a human-in-the-loop approach.
When synthesizing an \alltoall{} collective using some communication sketches, \sysname's
contiguity encoding may take more time in proving the optimality of a feasible solution.
We put a time limit of 30 minutes on the contiguity encoding in these cases.
The contiguity encoding for sketch \emph{ndv2-sk-1} reaches this timeout,
but a feasible solution was already found in 4min 14s.
We have also been able to synthesize an \allgather{} for 80 GPUs (10 \ndvtwo nodes) in under 8 minutes.

\section{Related Work}

The MPI standard provides a set of collective communication algorithms
that enable efficient distributed computations of interconnected
nodes~\cite{mpi}.  The HPC community has focused on the efficient
implementation of these MPI collective
algorithms~\cite{pjevsivac2007performance,
  thakur2005optimization} and demonstrated how to build optimized
algorithms for specific interconnects, like mesh, hypercube, or
fat-tree~\cite{scott1991efficient,bokhari1992complete,barnett1993global}.
In contrast to \sysname, these prior works assume homogeneous
interconnects and are often only focused on bandwidth
optimality. 
Hybrid algorithms~\cite{chan2007collective,barnett1993global} combine bandwidth- and latency-optimal algorithms based on input sizes, but only qfor mesh networks. 

NCCL~\cite{ncclrepo} is a GPU implementation of a subset of the standard MPI collectives, optimized for NVLINK and
Infiniband interconnects.  While NCCL uses the topology of GPU
connections and NIC placement along with buffer size to decide between
two main types of communication algorithms --- Ring and Tree, it is
agnostic to the exact performance profile of the links, and thus (as
we show) is often multiple times slower than \sysname's topology aware
collectives.

Recent works like SCCL~\cite{cai2021synthesizing}, Blink~\cite{wang2018blink}, and
Plink~\cite{plink} specialize algorithms for the underlying
topology. SCCL solves an integer programming encoding based on discrete-time values
in the form of steps and rounds of the algorithm in order to achieve the pareto-frontier of
latency- and bandwidth-optimal algorithms.
SCCL is able to synthesize a novel pareto-optimal \allgather algorithm for an Nvidia DGX1 node,
but its restrictive formulation constrains it to only only synthesize algorithms for single-node topologies.
\sysname on the other hand synthesizes collective algorithms for multi-node topologies.
Blink uses a heuristic spanning-tree packing algorithm to
maximize bandwidth utilization within a node and a hierarchical
approach across. Blink has good performance over NCCL in the case when NCCL cannot create rings spanning all GPUs inside a node.
\sysname, on the other hand, outperforms NCCL when using the entire node of GPUs.
Plink constructs a logical topology based on
bandwidth and latency probes of the physical topology to avoid
oversubscribed and congested links and searches for a reasonable
clustering of nodes for a two-level hierarchical reduction
strategy.  Plink builds that hierarchical
reduction from known primitives and does not search over the space of
possible algorithms.

There are also hierarchical approaches to implement
collectives~\cite{blueconnect,alex2018horovod,plink,wang2018blink}. For
example, Horovod~\cite{alex2018horovod} implements an \allreduce by a
local ReduceScatter, a global \allreduce, and then a local
\allgather. These methods do not search over possible
algorithms, but instead pick from a known set of decompositions.
Concurrent to our work, Ningning et al.~\cite{ningningMLSys} use syntax guided synthesis to 
combine base MPI primitives among a subset of nodes to hierarchically generate larger MPI primitives for the entire network. 
In contrast, \sysname uses a fine grained approach for algorithm synthesis while using communication sketches for scalability.   
Combining these two complementary approaches is an interesting opportunity for future work. 

Program sketching~\cite{solarlezama2008,jeon2015jsketch,10.1145/1168857.1168907} is a popular technique that has been applied to a variety of problems from 
synthesizing stencil computations~\cite{10.1145/1250734.1250754}, converting hand drawings to images~\cite{ellis2018learning} to social media recommendations~\cite{10.1145/2396761.2398507}. Our work builds on this body of work to use sketching to effectively search a large space of communication algorithms. 

Lastly, network flow problems have used linear programming to solve routing and scheduling
problems for traffic engineering~\cite{swan, b4, tempus, radwan, cascara} and
topology engineering~\cite{shoofly}.
These techniques, however, cannot be used for generating collective algorithms since
communication collectives do not follow all flow properties.
Non-source GPUs in a collective can send the same chunk over different links in parallel while
having received that chunk only once, which violates an important flow-conservation property
used extensively in network flow problem literature.
\sysname on the other hand makes use of communication sketches and an encoding relaxation technique to
solve a continuous-time integer linear programming that faithfully models communication collectives.

\section{Conclusion and Future Work}\label{sec:conclusion}
\sysname is a topology and input-size aware collective communication library
for multi-node distributed machine learning training and inference.
\sysname uses user-provided communication sketches to guide synthesis of collective algorithms.
Using a three-step technique of relaxed routing, heuristic ordering, and contiguity and exact scheduling,
\sysname generates efficient collectives for multi-node topologies.
We also make some brief observations about \sysname below:

\myparab{Scalability.} \sysname can synthesize algorithms for large-scale nodes -
we have been able to synthesize an \allgather algorithm for 8 Azure \ndvtwo nodes
using \sysname in under 5 minutes. As compared to NCCL, this algorithm has up-to $1.7\times$
higher algorithm bandwidth for different data sizes. We also evaluated \sysname's synthesis for
8 Nvidia \dgxtwo nodes (128 GPUs) and found a solution in around 11 hours.
While \sysname scales to multi-node topologies, the synthesis technique
is still based on solving an NP-hard problem that grows exponentially with a quadratic power
with scale. As a future work, we would like to scale \sysname further by hierarchically composing
synthesized algorithms.

\myparab{Generality across different topologies.} Apart from hierarchical topologies like
Nvidia \dgxtwo and Azure \ndvtwo, \sysname can also be applied to non-hierarchical topologies
like a 2D-Torus. We were able to synthesize an \allgather algorithm for a 2D $6\times 8$ Torus using
\sysname. We made use of the symmetry attribute in communication sketches to explore synthesis
for this topology.
However, the amount of exploration we can do with different communication sketches may be
more limited in these cases than for hierarchical topologies.

\myparab{Exploring communication sketches.} Communication sketches have proven effective in
narrowing the search space of algorithms. Interestingly,
different communication sketches can optimize different ranges of
input sizes. Communication sketches reflect the intuition of developers, and by intelligently
exploring the space of communication sketches we can obtain a range of collective algorithms with
different performance characteristics.
Learning an automated controller for exploring communication sketches is an interesting
direction for collective algorithm synthesis in the future.

To conclude, \sysname uses the abstraction of communication sketches and
a novel problem formulation to generate efficient algorithms for collectives like
\allgather, \alltoall, and \allreduce.
The algorithms thus generated are up-to $6.7\times$ faster than the state-of-the-art NCCL
and result in $11\% - 2.4\times$ faster end-to-end training time.
\section*{Acknowledgements}
We would like to thank our shepherd, Aurojit Panda, the anonymous reviewers at NSDI'23,
and the members of the Systems and Storage Lab at UT Austin for their insightful
comments and suggestions. This work was partially supported by NSF CAREER \#1751277, the UT
Austin-Portugal BigHPC project (POCI-01-0247-FEDER-045924), and donations from VMware.





\bibliographystyle{plain}
\bibliography{main}

\section*{Appendix}
\appendix
\camera{\section{Communication Sketch Input}\label{app:sketch}}
\sysname adopts a user-in-the-loop approach where algorithm
designers provide a communication sketch to guide communication
algorithm synthesis by \sysname.
\sysname's synthesizer takes in a profiled topology provided by \sysname profiler
along with a communication sketch provided by a human-in-the-loop.
A communication sketch comprises of a logical topology, switch-hyperedge strategy,
symmetry information, input size, and other hyperparameters.
Listing~\ref{list:dgx2-sk1} gives an example of how users can provide a
communication sketch input to the \sysname synthesizer. Here, we
show an example of the communication sketch \textit{dgx2-sk-1} used in the
evaluation to synthesize an \allgather algorithm for 2 Nvidia \dgxtwo nodes (each
node has 16 GPUs and 8 NICs, every two GPUs in the node share a NIC).

The sketch annotates the NVSwitch in each node and sets a \ucmin switch-hyperedge
strategy. Further, the inter-node sketch fixes the sender and receiver GPUs in a
node for inter-node data transfers. In our example, the odd-numbered GPUs sharing a NIC
are chosen as senders and the even-numbered GPUs are chosen as receivers for inter-node
communication.
The user also annotates how the inter-node relay GPUs would split the inter-node bandwidth
using a \textit{beta\_split} attribute.
Since only a single GPU per NIC is chosen in our example to perform inter-node send and
similarly receive, the bandwidth is not split.
Optionally, the user can also map chunks to sender GPUs so that only mapped GPUs are used for
inter-node transfers for the chunk.
The \textit{chunk\_to\_relay\_map} attribute defines the parameters for the mapping function.
The communication sketch also allows users to play with rotational symmetry
for data routing.
Given a symmetry offset and a group size, a chunk transfer over a link is set to be equivalent to
a rotationally symmetric chunk over a rotationally symmetric link. In our example of the
\textit{symmetry\_offset} attribute,
using $[2, 16]$ fixes an intra-node symmetry with an offset of two, and using
$[16, 32]$ fixes a symmetric data transfer pattern between the two \dgxtwo nodes.
Hyperparameters like input data partitioning and input size can also be provided via
the communication sketch.

\begin{lstlisting}[caption={Example sketch \textit{dgx2-sk-1} for \allgather},label={list:dgx2-sk1}]
{
    // sketch for intra-node policy
    "intranode_sketch": {
        "strategy": "switch",
        "switches": [[0,1,2,3,4,5,6,7,8,9,10,11,12,13,14,15]],
        "switch_hyperedge_strategy": ["uc-min"]
    },

    // sketch for communication policy between any two nodes
    "internode_sketch": {
        "strategy": "relay",
        "internode_conn": {"1" : [0], "3" : [2], "5" : [4], "7" : [6], "9" : [8], "11" : [10], "13" : [12], "15" : [14]}, // "i": [j1, j2] implies GPU i in a node will only send data to GPU j1 and j2 of another node
        "beta_split": {"1": 1, "3": 1, "5": 1, "7" : 1, "9" : 1, "11" : 1, "13" : 1, "15" : 1}, // "i": n implies inter-node sends from a GPU i of a node will use 1/n-th of the inter-node bandwidth
        "chunk_to_relay_map": [2,1] // maps chunk to a sender relay GPU. [r1,r2] means chunk c will be send to another node via GPU (rp//r1)*r1 + r2, where rp is the precondition GPU for chunk c
    },

    // enforces rotational symmetry.
    // [(o,g), ..]: o is the rotational offset and g is the group size for the rotational symmetry.
    // : eg. send(c,src,r) == send( (c + o)%g, (src + o)%g, (r + o)%g)
    "symmetry_offsets": [[2, 16], [16, 32]],

    "hyperparameters": {
        "input_chunkup": 2, // Data at each GPU is partitioned into 2 chunks that can be independently routed
        "input_size": "1M"
    }
}
\end{lstlisting}

\section{\sysname Synthesizer in Detail}\label{app:encoding}
As explained in Section~\ref{sec:distsccl-synthesizer}, \sysname's synthesizer has
routing, heuristic ordering, and contiguity and exact scheduling stages. We provide
a detailed description of each of these stages in this section. We first formally
introduce some terms that we will use later. Let $\mathcal{C}$ denote the set of
chunks that are required to be routed in the algorithm for collective $coll$. Let
$\mathcal{R}$ denote the set of GPU ranks involved in $coll$. Let $coll\text{.precondition}$
and $coll\text{.postcondition}$ denote the precondition and post-condition of the collective
respectively.The tuple
$(c,r) \in coll\text{.precondition}, c \in \mathcal{C}, r \in \mathcal{R}$,
if chunk $c$ is present at rank $r$ at the start of the collective.
Similarly, the $(c,r) \in coll\text{.postcondition}$
if chunk $c$ has to be present at rank $r$ at the end of the collective.
Further, let $\mathcal{L}$ denote the set of links,
such that $(r1,r2) \in \mathcal{L}, r1 \in \mathcal{R}, r2 \in \mathcal{R}$ if
there exists a link from rank $r1$ to rank $r2$ in the logical topology determined by the
topology and communication sketch.
Let $\mathcal{S}^{send}_r$ denote the set of switched destinations for rank $r$, such that
$dst \in \mathcal{S}^{send}_r$ if link $(r,dst)$ is a part of a switch-hyperedge.
Similarly, $\mathcal{S}^{recv}_r$ denotes the set of switched sources for rank $r$, such that
$src \in \mathcal{S}^{recv}_r$ if link $(src,r)$ is a part of a switch-hyperedge.
$\alpha(r1,r2)$, $\beta(r1,r2)$ are the alpha and beta costs
respectively of the link $(r1,r2) \in \mathcal{L}$.
The term $lat(r1,r2)$ is the sum of $\alpha(r1,r2)$ and $\beta(r1,r2)$ cost of the link, which denotes the
total transfer cost of a single chunk over link $(r1,r2)$.
Table~\ref{tab:milp-variables} lists the variables that the \sysname's synthesizer solves for.
We will describe each variable in detail in this section.

\subsection{Routing}\label{app:pathenc}
    The main aim of the routing stage is to give us the path that every chunk takes in the collective.
    Our objective is to minimize the time (denoted by continuous variable $time$) it takes to reach the
    post-condition of the collective.
    \begin{align}
        \text{Minimize} \quad time \label{cstr:obj}
    \end{align}    
    
   The time taken for the collective algorithm is the latest time at which a chunk becomes
    available on a rank that is in the post-condition of the collective.  We use a continuous
    variable $start[c,r]$ to denote the time that chunk $c$ becomes available on rank $r$, and
    end up with the following constraints for $time$
    \begin{align}
        time \geq start[c,r] \quad \forall (c,r) \in coll\text{.postcondition}
    \end{align}

    For chunks on ranks that belong to the collective's precondition, we set the start time to zero.
    \begin{align}
        start[c,r] = 0 \quad \forall (c,r) \in coll\text{.precondition}
    \end{align}

    We also add correctness constraints in our formulation for routing - chunks are sent from a
    GPU rank only after they have been received on that rank. We introduce a continuous variable
    $send[c,src,r]$ to denote the time of sending chunk $c$ from rank $src$ to rank $r$ and add
    the following constraint to our formulation:
    \begin{align}
        send[c,src,r] \geq start[c,src] \quad \forall c \in \mathcal{C} \quad \forall (src,r) \in \mathcal{L} \label{cstr:corr}
    \end{align}

    We use a binary variable $is\_sent[c,src,r]$ to indicate if chunk $c$ is sent over the link
    $(src,r)$ in our algorithm.
    We note that the routing stage does not strictly respect bandwidth constraints of any link - the
    generated solution may send two chunks simultaneously over a link at the time cost of one chunk.
    The chunk start time on a rank will be determined only by the chunk send time on the source,
    independent of other chunk transfers on the link (eq.~\ref{relbw}).
    LHS$\rightarrow$RHS in the equation signifies an indicator constraint, i.e., if LHS is 1, RHS will hold.
    \begin{equation}
        \begin{split}
        is\_sent[c,src,r]\!\rightarrow\!start[c,r] & = send[c,src,r] + lat(src,r) \\
        & \forall c \in \mathcal{C} \quad \forall (src,r) \in \mathcal{L} \label{relbw}\\
        \end{split}
    \end{equation}
    Instead of bandwidth constraints, this encoding uses \emph{relaxed bandwidth constraints}.
    They are expressed by aggregating the link transfer time of all chunks sent over a link and
    using it to to lower bound the total time of the algorithm (eq.~\ref{link1}).
    For switched connections, the total time is lower bounded by the sum of link transfer times of
    all chunks sent over all switched outgoing links from a source, and also by the sum of link transfer
    times for chunks received from all incoming links to a destination (eq.~\ref{switch1s1} and eq.~\ref{switch1r1}).
    \begin{align}
        time \geq \sum_{c \in C} (lat(src,r) * is\_sent[c,src,r]) \quad \forall (src,r) \in \mathcal{L} \label{link1}\\
        time \geq \sum_{c \in C} \sum_{dst \in \mathcal{S}^{send}_r} (lat(r,dst) * is\_sent[c,r,dst]) \quad \forall r \in \mathcal{S}_{send} \label{switch1s1}\\
        time \geq \sum_{c \in C} \sum_{src \in \mathcal{S}^{recv}_r} (lat(src,r) * is\_sent[c,src,r])  \quad \forall r \in \mathcal{S}_{recv}\label{switch1r1}
    \end{align}

    Based on the communication sketch, we also add constraints for $\ucmax$ and $\ucmin$ strategies for switch-hyperedges
    to maximize and minimize the number of links utilized in a switch respectively.
    We introduce a new binary variable $is\_util[src,r]$ for links $(src,r)$ that are a part of a switch-hyperedge.
    This variable is $1$ if any chunk is sent over link $(src,r)$, and $0$ otherwise.(eq.~\ref{eq:and1} and eq.~\ref{eq:and2}).
    According to the switch-hyperedge strategy, we add this variable, weighted with a small constant $\gamma$, to the objective function (eq.~\ref{eq:uc}).
    $\gamma$ is negative for $\ucmax$ and positive for $\ucmin$.

    \begin{align}
        is\_util[src,r] >= is\_sent[c,src,r] \quad \forall{c \in \mathcal{C}} \forall{(src,r) \in \mathcal{L}} \label{eq:and1}
    \end{align}

    \begin{align}
        is\_util[src,r] <= \sum_{\forall{c \in \mathcal{C}}} is\_sent[c,src,r]\quad \forall{(src,r) \in \mathcal{L}} \label{eq:and2}
    \end{align}

    \begin{align}
        \text{Minimize} \quad time + \gamma \times (\sum_{(src,r): \text{switched links}} is\_util[src,r]) \label{eq:uc}
    \end{align}

    We also add symmetry constraints according to the symmetry offsets provided by user in the communication sketch.
    For a chunk $c$ and link $(src,r)$, we identify a rotationally symmetric chunk $\hat{c}$ and
    link $(\hat{src},\hat{r})$ and add the following constraints:
    \begin{align}
        start[c,r] = start[\hat{c},\hat{r}] \\
        send[c,src,r] = send[\hat{c},\hat{src},\hat{r}] \\
        is\_sent[c,src,r] = is\_sent[\hat{c},\hat{src},\hat{r}]
    \end{align}
    
    Further, for chunks that start on one node and have a final destination on another node,
    we add inter-node transfer constraints which specify that at least one inter-node link will be used to transfer that chunk.
    \begin{align}
        \sum_{(r_1,r_2) \in \mathcal{L} : r_1 \in \text{node}_1, r_2 \in \text{node}_2} is\_sent[c,r_1,r_2] \geq 1
    \end{align}

\begin{table}[t]
    \tablestyle{5pt}{1.12}\begin{tabular}{@{}lp{180pt}@{}}
      MILP Variables & Explanation \\
      \hline
      \textbf{Routing} & \\
      $time$        & time spent in the collective algorithm \\
      $start[c,r]$    & time at which chunk $c$ becomes available at GPU $r$      \\
      $send[c,src,r]$ & time at which chunk $c$ is sent from GPU $src$ to GPU $r$ \\
      $is\_sent[c,src,r]$ & indicates if chunk $c$ is sent from GPU $src$ to GPU $r$ \\
      $is\_util[src,r]$ & indicates if any chunk is sent from GPU $src$ to GPU $r$ \\
      & \\
      \textbf{Contiguity}  & \\
      $is\_together[c,o,r]$ & indicates if chunks $c$ and $o$ are sent to GPU $r$ together from the same source, thus sharing the bandwidth and reducing the latency cost of transfer \\
    \end{tabular}
    \caption{Variables used in \sysname's MILP formulation.
    Variables with prefix \textit{is\_} are binary variables and
    others are continuous variables.}
    \label{tab:milp-variables}
  \end{table}

\subsection{Ordering Heuristics}\label{app:ordheur}
    We start the heuristic ordering by determining the paths each chunk takes using the solution of the path encoding.
    We then consider the first link in every path as a candidate for scheduling a chunk transfer.
    Using heuristics like \textit{chunk-with-shortest-path-until-now-first} and
    \textit{chunk-with-longest-path-from-now-first}, we select a path (and thus a chunk) which should be scheduled in this round.
    We keep a running estimate of link time, which is the earliest time at which a chunk can be scheduled over the link.
    We also keep a running estimate of chunk time, which is the earliest time at which
    the next link transfer can be scheduled for a chunk.
    At the start, the link time for every link is 0 and the chunk time for every chunk is 0.
    When a path is chosen in the first round, the chunk associated with the path is scheduled to traverse the first link in the path.
    The link time of that link increases by link latency and chunk time of that chunk increases by link latency.
    The link candidate from the selected path is also updated to be the next link in the path.
    For the next rounds, we decide which path's candidate link to schedule next using the tracked link and chunk times along with the scheduling heuristics.
    This keeps going until we have scheduled a data transfer over all the links in all the paths.
    We find that the best heuristics differ for architectures with NVLinks and those with NVSwitches,
    in terms of whether to start selecting links to schedule in the same order as the paths or in the opposite order of the paths.
    The heuristic ordering has the following three outputs:
    \begin{itemize}
        \item $\text{chunk\_order}(r_1,r_2)$, an ordered list of chunks transferred along each link $(r_1,r_2)$. If chunk $c_1$ is present
        before chunk $c_2$ in $\text{chunk\_order}(r_1,r_2)$, it denotes that $c_1$ is scheduled to be sent before $c_2$ over
        link $(r_1,r_2$).
        \item $\text{switch\_send\_order}(r)$, an ordering on the chunks sent from a switch source $r$ to any of the switch destinations
        $dsts$. If $(c_1,dst_1)$ is present before tuple $(c_2,dst_2)$ in $\text{switch\_send\_order}(r)$, it means that a send of $c_1$
        over link $(r,dst_1)$ should be scheduled before a send of chunk $c_2$ over link $(r,dst_2)$.
        \item $\text{switch\_recv\_order}(r)$, an ordering on the chunks received on a switch destination $r$ from any of the switch sources
        $srcs$. If $(c_1,src_1)$ is present before tuple $(c_2,src_2)$ in $\text{switch\_recv\_order}(r)$, it means that a receive of $c_1$
        over link $(src_1,r)$ should be scheduled before a receive of chunk $c_2$ over link $(src_2,r)$.
    \end{itemize}

\subsection{Contiguity and Exact Scheduling}\label{app:contenc}
Finally, we describe the formulation for the contiguity and exact scheduling stage.
Given the link and switch ordering from the heuristic ordering stage,
the aim of this stage is to find the sweet spot in the trade-off between
lower link latency by sending multiple data chunks contiguously as a big
data chunk and reduced pipelining benefits due to the big data-chunk transfer.
We provide the main set of constraints in our formulation below, leaving out
other less important constraints.

Our objective is still to minimize the time of the collective and constraints
eq.~\ref{cstr:obj}-eq.~\ref{cstr:corr} must still hold in this formulation.
We add a new binary variable $is\_together(c_1,c_2,r)$ for all chunks $c_1$ and $c_2$
that are sent over the same link to rank $r$. If $is\_together(c_1,c_2,r)$ is $1$,
chunks $c_1$ and $c_2$ are sent as a single data-chunk over a link to rank $r$.
\begin{equation}
    \begin{split}
    is\_together[c,o,r]\!\rightarrow\!send[c,src,r] = send[o,src,r] \\
    \forall{c,o \in \text{chunk\_order}(src,r)} \quad \forall{(src,r) \in \mathcal{L}}
    \end{split}
\end{equation}
The transfer time of a data chunk $c$ along a link $(src,r)$ will be determined by all other data chunks
that it has to travel together with:
\begin{equation}
    \begin{split}
    lat[c,src,r] =& \alpha(src,r) + \beta(src,r) * \\
    &(\sum_{o \in \text{chunk\_order}(src,r)} is\_together[c,o,r])\\
    &\forall{c \in \text{chunk\_order}(src,r)} \quad \forall{(src,r) \in \mathcal{L}}\\
    \end{split}
\end{equation}
\begin{equation}
    \begin{split}
    start[c,r] =& send[c,src,r] + lat[c,src,r]\\
    &\forall{c \in \text{chunk\_order}(src,r)} \quad \forall{(src,r) \in \mathcal(L)}\\
    \end{split}
\end{equation}

We also add strict bandwidth constraints for this formulation, allowing only one data chunk per link transfer time
if the data chunks are not sent contiguously over the link.
Let $pos(c,src,r)$ determine the position of chunk $c$ in the $\text{chunk\_order}(src,r)$, then
\begin{equation}
    \begin{split}
    \neg is\_together[c,o,r]\!\rightarrow\!send[o,src,r]\geq\!send[c,src,r]\\
    +lat[c,src,r]
    \quad \forall{c \in \text{chunk\_order}(src,r)}\\
    \forall{o \in \text{chunk\_order}(src,r)} \\
    \text{if} \quad \text{pos}(o,src,r) \ge \text{pos}(c,src,r)
    \quad \forall{(src,r) \in \mathcal{L}}
    \end{split}
\end{equation}

Similarly, we add bandwidth constraints for switch, allowing a source to send data to only one switched destination at a time,
and a receiver to receive data from only one switched sender at a time.
Let $sw-pos-send(c,r,dst)$ determine the position of tuple $(c,dst)$ in the $\text{switch\_send\_order}(r)$,
and let $sw-pos-recv(c,src,r)$ determine the position of tuple $(c,src)$ in the $\text{switch\_recv\_order}(r)$, then,
\begin{equation}
    \begin{split}
    send[o,r,dst_o]\geq\!send[c,r,dst_c]+lat[c,r,dst_c]\\
    \forall{(c,dst_c) \in \text{switch\_send\_order}(r)}\\
    \forall{(o,dst_o) \in \text{switch\_send\_order}(r)}\\
    \text{if} \quad \text{sw-pos-send}(o,r,dst_o) \ge \text{sw-pos-send}(c,r,dst_c)\\
    \quad \forall{r \in \mathcal{S}^{send}}
    \end{split}
\end{equation}
\begin{equation}
    \begin{split}
    send[o,src_o,r]\geq\!send[c,src_c,r]+lat[c,src_c,r]\\
    \forall{(c,src_c) \in \text{switch\_recv\_order}(r)}\\
    \forall{(o,src_o) \in \text{switch\_recv\_order}(r)}\\
    \text{if} \quad \text{sw-pos-recv}(o,src_o,r) \ge \text{sw-pos-recv}(c,src_c,r)\\
    \quad \forall{r \in \mathcal{S}^{recv}}
    \end{split}
\end{equation}

\section{Standalone Experiments on Four Azure \ndvtwo Nodes}
\label{app:four-node}

\begin{figure}[!tb]
  \centering
    \includegraphics[width=\linewidth]{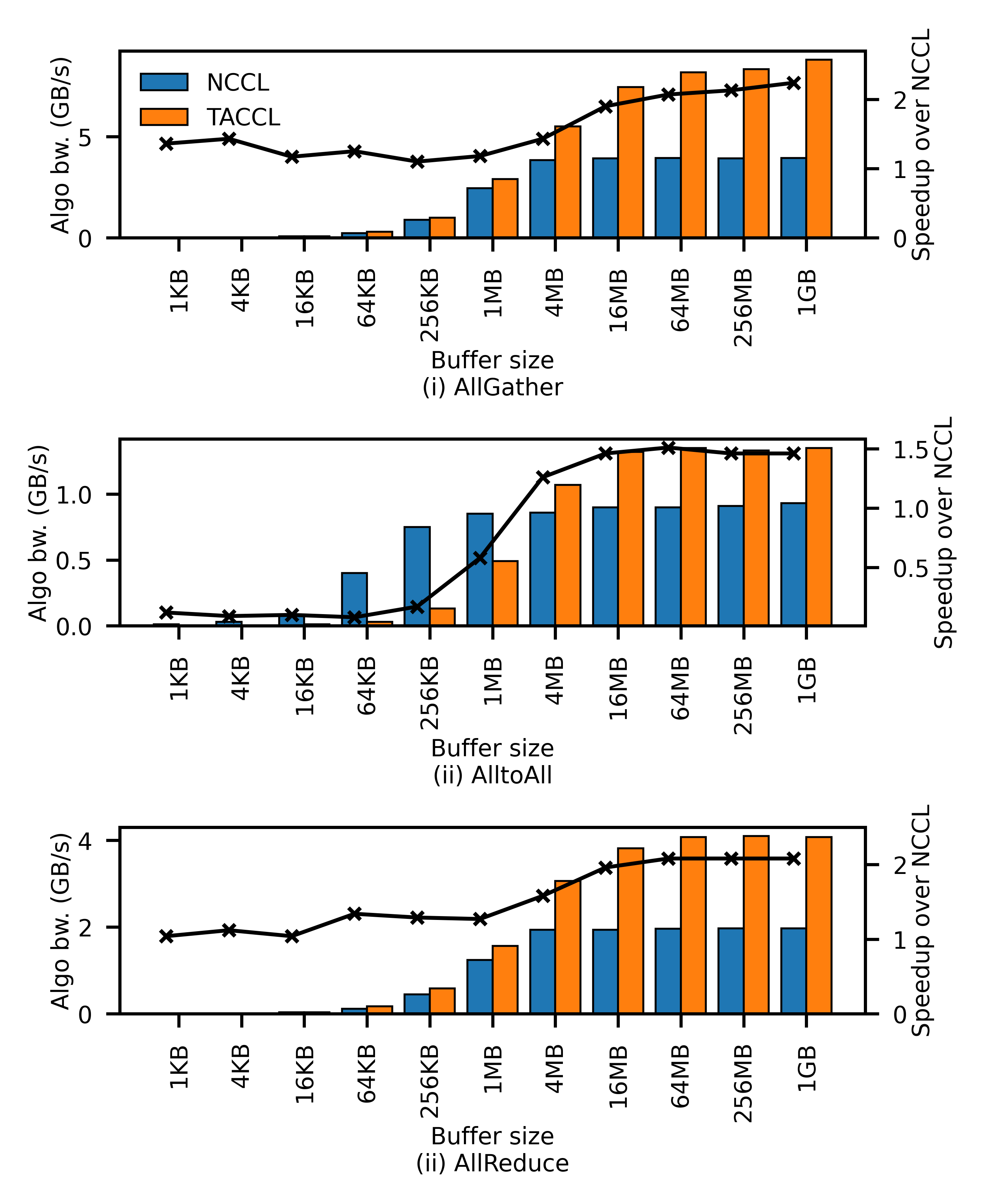}
    \caption{\small{Algorithm bandwidth of \sysname algorithms compared against NCCL (left Y-axis)
    and their speedup over NCCL (right Y-axis) for \allgather, \alltoall, and \allreduce
    collectives on four \ndvtwo nodes.}}
    \label{fig:ndv2-app}
  \end{figure}

Figure~\ref{fig:ndv2-app} shows additional algorithm bandwidth and the speedup
over NCCL graphs of \sysname for \allgather, \alltoall, and \allreduce on 4-node
\ndvtwo{} cluster. We synthesize all collectives using the \emph{ndv2-sk-1}
communication sketch (see Section~\ref{sec:standalone} for details), and
lower them using 1 or 8 instances. We plot the best of the two algorithms over
different buffer sizes.

\sysname's \allgather algorithms are $10\% - 2.2\times$ faster than NCCL across
all buffer sizes.
For \alltoall, the \emph{ndv2-sk-1} sketch is most effective for large buffer sizes,
and helps generate algorithms that are up-to $46\%$ faster than NCCL for buffer
size greater than $1$MB.
\sysname \allreduce algorithms are up-to $34\%$ faster than NCCL
for small buffer sizes and $1.9\times - 2.1\times$ faster than NCCL for
larger buffer sizes.

\end{document}